\RequirePackage{fix-cm}
\documentclass[smallextended]{svjour3}       
\smartqed  
\usepackage{scrtime}
\usepackage[dont-mess-around]{fnpct}

\usepackage{etoolbox}
\usepackage{graphicx}
\usepackage[utf8]{inputenc}
\usepackage{amsfonts}

\usepackage{amsmath}
\usepackage{amssymb}

\usepackage{mathrsfs}
\usepackage{xcolor}
\usepackage{float}
\usepackage{bbm}
\usepackage{todonotes}
\usepackage{soul}
\usepackage[titletoc]{appendix}
\usepackage{algorithm,algorithmic}
\usepackage{cancel}

\usepackage{fancyhdr}
\usepackage{url}
\usepackage{comment}
\usepackage{multirow}
\usepackage[bottom]{footmisc}
\usepackage{subfigure}
\usepackage{booktabs}
\usepackage{enumitem}
\usepackage{framed}
\usepackage{caption}

\newcommand*{\affaddr}[1]{#1} 
\newcommand*{\affmark}[1][*]{\textsuperscript{#1}}
\newcommand{\cla}[1]{\textcolor[rgb]{0,0,0}{#1}}

\newcommand{\col}[1]{\textcolor[rgb]{0,0,0}{#1}}

\newcommand{\R}{\mathbb{R}}

\newcommand{\N}{\mathbb{N}}

\newcommand{\pp}{\mathbb{P}}

\newcommand{\kR}{\mathcal{R}}

\newcommand{\kF}{\mathcal{F}}
\newcommand{\kG}{\mathcal{G}}

\newcommand{\kH}{\mathcal{H}}

\newcommand{\kN}{\mathcal{N}}
\newcommand{\kX}{\mathcal{X}}

\newcommand{\n}{n}

\newcommand{\lin}{\left[\kern-0.15em\left[}
\newcommand{\rin} {\right]\kern-0.15em\right]}
\newcommand{\linf}{[\kern-0.15em [}
\newcommand{\rinf} {]\kern-0.15em ]}
\newcommand{\ilin}{\left]\kern-0.15em\left]}
\newcommand{\irin} {\right[\kern-0.15em\right[}

\newcommand{\eqan}[1]{\begin{align}#1\end{align}}

\newcommand{\be}{\begin{equation}}
\newcommand{\ee}{\end{equation}}
\newcommand{\ER}{Erd\"os-Rényi\ }
\newcommand{\gb}{\beta }
\newcommand{\e}{{\rm e}}
\newcommand{\ds}{\displaystyle }
\newcommand{\za}{\alpha}

\newcommand{\vol} {\textbf}

\newcommand{\fpp}{\ln \left ( \frac{p_2(1-p)}{p(1-p_2)} \right)  }

\begin{document}

\title{{Approximating the cumulant generating function of triangles in the \ER random graph}
}


\author{Cristian Giardin\`a\protect\affmark[1]    \and
        Claudio Giberti\protect\affmark[2] \and \\
        Elena Magnanini\protect\affmark[1,]\protect\affmark[3]
}


\institute{
C.Giardin\`a \at
              \email{cristian.giardina@unimore.it}           
           \and
           C.Giberti \at
              \email{claudio.giberti@unimore.it}              
              \and
               E.Magnanini \at
              \email{elenam@math.unipd.it}\\\\              
\affaddr{\affmark[1] University of Modena and Reggio Emilia, via G.Campi 213/b, 41125 Modena, Italy.}\\
\affaddr{\affmark[2] University of Modena and Reggio Emilia, via G.Amendola 2, 42122 Reggio Emilia, Italy.}\\
\affaddr{\affmark[3] University of Padova, via Trieste 63, 35121 Padova, Italy.}                   
}

\date{Received: date / Accepted: date}

\maketitle

\begin{abstract}
We study the pressure of the ``edge-triangle model'', which is equivalent to the cumulant generating function of triangles in the \ER random graph. The investigation involves a population dynamics method on finite graphs of increasing volume, as well as a discretization of the graphon variational problem arising in the infinite volume limit.  As a result, we locate a curve in the parameter space where a one-step replica symmetry breaking transition occurs. Sampling a large graph in the broken symmetry phase is well described by a graphon with a structure very close to the one of an equi-bipartite graph. 
\end{abstract}

\section{Introduction}
\label{intro}
Sampling random graphs with prescribed macroscopic properties (such as a given density of certain subgraphs)
received considerable attention in recent years.
From a statistical physics perspective, one can think of two procedures:
\begin{itemize}
\item the {\em micro-canonical ensemble}, where the sampling
is performed with a uniform distribution over the set of all graphs that satisfy the macroscopic constraint exactly;
\item the {\em canonical ensemble}, where the sampling is 
done with respect to a larger set of graphs that satisfies the macroscopic constraint only on average.
\end{itemize}
\cla{We shall discuss here the simplest non trivial case, i.e. the constraint is on the number of edges and 
the number of triangles in the graph. In the micro-canonical ensemble these numbers are prescribed exactly. The canonical ensemble
is instead provided by the so-called {\em edge-triangle model}, which is defined by the Boltzamnn-Gibbs
distribution in which one tunes the average density of edges and the 
average density of triangles by varying the corresponding conjugate parameters}. 
The edge-triangle model is in turn the  simplest example of the more general 
{\em exponential random graph} class of models, in which one introduces several
parameters to control the density of an arbitrary set of subgraphs.

Whereas the equivalence in the thermodynamic limit of micro-canonical and canonical ensemble is  
true for several physical systems of interest (often the system is then studied in the canonical
ensemble, that is usually more analytically tractable than the micro-canonical one), for \cla{random graphs}
it has been shown that such equivalence can not be taken for granted. 
In particular \cite{den2018ensemble}
identified a region of values for the densities of edges and triangles where there is
{\em ensemble inequivalence}, as measured by  a positive relative entropy between the micro-canonical
and canonical measures. In this paper we address the following problem: 
\begin{quote}
How does a large graph look like when it is sampled from the edge-triangle model
(i.e. 
imposing some given {\em average values} for the densities of edges and triangle)?
Does the sampling from the edge-triangle
model give the same result of the sampling with respect to the microcanonical 
ensemble (
i.e. imposing  some given {\em exact values} of edge and triangle densities)?
\end{quote}

\subsection{The edge-triangle model and the \ER random graph}

To define the setting, let us  consider a {\em graph} with $\n$ vertices, that we identify with the elements of the set  $[\n]=\{1,2,3,\ldots, \n\}$. 
We shall describe the graph using  its adjacency matrix ${x}=(x_{i,j})_{i,j \in [\n]}$, defined as follows: the entry $x_{i,j}=1$ if the edge connecting vertex $i$ with vertex $j$ is present,  and $x_{i,j}=0$ otherwise. Since the graphs considered in this paper are undirected and without loops, the adjacency matrices will be always symmetric, with $0$ or $1$ entries and zeros on the diagonal. 
We will denote by $\kX_{\n}$ the set of adjacency matrices of size $n$, and we also define for later use $\kX = \cup_{n\ge2} \kX_{\n}$
the set of adjacency matrices \cla{of all sizes}.   
The number of edges and triangles in a graph represented by a matrix $x \in \kX_{\n}$ is given by
\be\label{eq:num-triang}
E_n(x)=   \sum_{1 \le i < j \le \n} x_{i,j}  \qquad\qquad\qquad T_n(x)= \sum_{1\le i < j < k \le n} x_{i,j}x_{j,k}x_{k,i}.
\ee
The previous quantities result in random variables if the graph is a {\em random graph}, i.e. if it is sampled from the set of $2^{{\n \choose 2}}$ undirected simple graphs 
with $\n$ vertices according to some probability distribution.  
The  simplest possible distribution is the so-called \col{{\em \ER model}}, where  each pair of vertices is connected with probability  $p>0$, independently of the other pairs. Thus, in the \ER case  the entries of the adjacency matrix, ${x}=(x_{i,j})_{i,j \in [\n]}$, form a set of  independent  identically distributed (i.i.d.) Bernoulli variables with $\pp (x_{i,j}=1)=p$. 
In spite of the simple probabilistic set-up 
the large deviation principle for the \ER graph 
is far from simple and has been developed only in recent years \cite{chatterjee2010applications,chatterjee2011large,chatterjee2016nonlinear,lubetzky2015replica,chatterjee2017large,zhao2017lower,dembo2018large,lopez2020imaginary,lopez2020transitions,metz2019condensation}.
In particular, it has been found that the large deviation function 
may be non-convex.

The exponential random graph model
is devised to enhance or decrease the probability of specific geometric structures in the random graph. 
Here we \cla{define} the \col{{\em edge-triangle model}} that involves only triangles and edges \cite{Newman}. Let $\gb_1, \gb_2 \in \R$, then the probability of a graph with
adjacency matrix  $x\in \kX_{n}$ \cla{in the edge-triangle model} is given by:
{
\be\label{eq:prob-exprg}
\nu_\n (x)=\frac{\exp \left ( \frac{\ds 6\beta_2}{\ds n} \ds T_n(x) +  2 \beta_1 \ds E_n(x) \right ) } {Z_\n(\beta_1,\beta_2)}, 
\ee
}
where 
$Z_\n(\beta_1,\beta_2)$ is the {\em partition function}, i.e.  the normalizing factor
\be\label{eq:partit-expon}
Z_\n(\beta_1,\beta_2)=\sum_{x \in \kX_{n}} \exp \left ( \frac{\ds 6 \beta_2}{\ds n} \ds T_n(x) +  2 \beta_1 \ds E_n(x) \right ) .
\ee
\cla{The factors $6$ and $2$ are conventional in the definition and account for the permutations of 3 vertices of a triangle
and the 2 vertices of an edge.}
The \ER model with paramemeter $0<p<1$ is embedded in the edge-triangle model, since its distribution: 
\be\label{eq:prob-ee} 
\nu_n^{ER} (x) =\prod_{1 \le i < j \le n} p^{x_{i,j}} (1-p)^{1-x_{i,j}} = (1-p)^{n \choose 2} \e^{ h_p E_n(x)}, \qquad \mbox{where}\quad h_p:= \log \frac{p}{1-p},
\ee
can be obtained from \eqref{eq:prob-exprg} by setting {$\beta_1=h_p/2$} and $\beta_2=0$. 
If $\beta_2>0$  the probability of finding triangles is enhanced with respect to the \ER case, while it is decreased if  $\beta_2<0$.  
In the limiting case $\beta_2\to -\infty$ the edge-triangle model \eqref{eq:prob-exprg}  gives zero probability to graphs containing triangles.

A key quantity in the study of the thermodynamic properties is the {\em pressure},
that at finite volume is defined as
\be\label{eq:press-exprg}
\psi_n(\beta_1,\beta_2) = \frac{1}{n^2}  \ln Z_\n(\beta_1,\beta_2).
\ee 
By taking derivatives with respect to the model parameters one computes the averages
of the edge density  \cla{$e=2 E_n/\n^2$} and of the triangle density \cla{$t=6\,T_n/\n^3$}:
\be
\label{averages}
\cla{\langle e\rangle_n = \frac{\partial \psi_n}{\partial \beta_1},  \qquad\qquad
\langle t \rangle_n = \frac{\partial \psi_n}{\partial \beta_2}},
\ee
where \cla{$\langle\cdot\rangle_n$} denotes expectation w.r.t. the measure $\nu_n$ {defined in} \eqref{eq:prob-exprg}.
We shall be interested in the behavior of very large graphs which mathematically 
is described by the (thermodynamic) limit $\n\to\infty$. General convexity arguments
imply that the thermodynamic limit is well defined, so that the infinite volume
pressure exists
\be\label{eq:press-exprg}
\psi(\beta_1,\beta_2) := \lim_{\n \to \infty} \frac{1}{n^2}  \ln Z_\n(\beta_1,\beta_2),
\ee 
and by Lebesgue dominated convergence limits and derivatives can be interchanged so that
the relation \eqref{averages} gives in the thermodynamic limit
\be
\label{averages2}
\langle e \rangle = \frac{\partial \psi}{\partial \beta_1}  \qquad\qquad
\langle t \rangle = \frac{\partial \psi}{\partial \beta_2},
\ee
where 
\cla{$\langle e \rangle = \lim_{n\to\infty} \langle e \rangle_n$} and $\langle t \rangle = \lim_{n\to\infty} \langle t \rangle_n$.
We shall work with the parametrization $(\beta_{1},\beta_{2})=(h_{p}/2,\alpha/6)$ where we recall $h_{p}=\ln\frac{p}{1-p}$ and $0<p<1$.
In this way the pressure of the edge-triangle model can be read as the cumulant generating function of the number of triangles in the  
\ER random graph. In other words, defining 
\be\label{eq:cumg}
\mu_{\n,p}(\za): = \frac{1}{ {\n \choose 2}} \log \left \langle \exp \left (\frac{\za}{n}T_n(X) \right ) \right \rangle^{ER}_n, \quad \za \in \R,
\ee  
where $  \langle  \cdot  \rangle^{ER}_n$ denotes the expectation w.r.t. the measure $\nu_n^{ER} $,
by a simple computation one can show that
\begin{align}\label{eq:funzg-pressure}
\hspace{-1cm}
\mu_{n,p}(\alpha)&= \ln(1-p)  +2\psi_n\left (\frac{h_p}{2} ,\frac{\alpha}{6}\right).
\end{align}
Thus, studying the cumulant generating function of the number of triangles in the  
\ER random graph is equivalent to studying the pressure of the edge-triangle model. 
%
%

\cla{To the best of our knowledge, the sampling from the canonical ensemble
has been investigated only in a limited region of the parameters ($\beta_1,\beta_2$),
see the review of known results in Section \ref{known}.
It is the aim of this paper to conduct a systematic exploration 
of the full parameter space by means of numerical simulations.
}

\subsection{Main results and paper organization}
\cla{We investigate the sampling from the canonical ensemble}
by studying the pressure of the edge-triangle model, or equivalently the 
cumulant generating function $\mu_p(\alpha)$ of triangles
in the \ER model with parameter $p$. \cla{We perform numerical simulations
 for finite graphs and compare them to the variational formulation
describing the infinite volume}. \col{
Our main results are the following:
\begin{itemize}
\item[$\bullet$] We shall collect multiple
evidences that the structure of graphs in the canonical ensemble has 
only two possibilities: it is either the constant graphon describing 
the \ER graph (i.e. independent edges, yet with a modified parameter for the probability of 
edges accounting for the imposed number of triangles) 
or it is the graphon describing the 1-step replica symmetric breaking solution 
(generalizing the bipartite random graphs that is know to be the exact solution for $ \alpha \to -\infty$). \\
\item[$\bullet$] By means of different numerical analysis (``cloning'' method for a direct measurement and ``gradient'' method
for the solution of the pressure variational problem) we shall identify 
a curve $\alpha_c(p)$ in the plane $(p,\alpha)$
separating these two regimes called, respectively, the replica symmetric phase 
and the replica symmetry broken phase. \\
\item[$\bullet$] We do not have a proof that replica symmetry 
broken phase $\alpha < \alpha_c(p)$ is entirely described by the 1-step replica symmetric breaking solution.
However, in contrast to the microcanonical sampling, our numerical analysis suggests that 
in the description of the canonical sampling no higher level of replica symmetry breaking is required.
As a consequence of the numerical analysis the value $\alpha_c(p)$ may be identified as a bona-fide 
critical value for a 1-step replica symmetry breaking transition.
\end{itemize}
}

\col{A remark is worth here on why we call the ``replica symmetry broken phase" the region of graph parameters where the solution is not homogeneous over the graph. Although replicas do not appear in our analysis,
we shall see that the optimizers of the variation problem yielding the solution have the typical  
block-structure of the ``overlap matrix'' in the replica symmetry broken phase of spin glasses.
}

\bigskip

The paper is structured as follows:
\begin{itemize}
\item In Section \ref{sectKnownRes} we review the variational formulation of the 
pressure. We recall the results that are known in the  literature
for the solution of the variational problem and 
\cla{describe} the 1-step replica symmetry breaking solution. 
\item In Section \ref{sectCloningRG} we present the numerical analysis of the
\cla{finite volume} pressure based on the ``cloning method'', which is a population dynamics algorithm.
\item In Section \ref{numsolxx} we solve a discretized version of the variational problem
in the infinite volume by a gradient projection method. \cla{Here we do not fix a-priori a specific structure for
the optimal graphon.}
\item In Section \ref{sez1stepRB} we solve the variational problem
restricted to a specific class of graphons, those  
corresponding to the 1-step replica symmetry breaking solution.
\end{itemize}
\cla{We will argue that the results of Sections \ref{numsolxx} and \ref{sez1stepRB} 
coincide (within numeral accuracy) and they are well approximated by the 
finite volume direct measurements of Section \ref{sectCloningRG}.
\col{Finally, Section \ref{per} contains the conclusion and some perspectives 
of this work on the probelm 
of ensemble inequivalence.}}
\section{Variational formulation}\label{sectKnownRes}
\subsection{Review of known results}
\label{known}

The theory of graph limits \cite{lovasz2006limits,lovasz2012large,borgs2006counting,borgs2008convergent,borgs2012convergent} relies on the  notion of {\em graphon},
which describes a random graph in the limit $n\to \infty$. A graphon is defined as a bounded Borel measurable function $f:[0,1]^{2}\longmapsto [0,1]$ that satisfies $f(x, y) = f(y,x)$ for all $x, y \in [0, 1]$. The idea behind this definition is a mapping of a  graph to the unitary square: intuitively the interval $[0,1]$ represents a continuum of vertices and  $f(x,y)$ is associated to the probability of 
connecting with an edge two vertices $x$ and $y$. For example, the graphon  describing  the   Erd\"os-Rényi random graph with parameter $p$ is the constant function $f$ identically equal to $p$. The set of graphons is denoted by $\mathcal{W}$. On this set  an  equivalence relation is introduced according to which 
$f, g \in \mathcal{W}$ are equivalent if there exists a bijection $\sigma :[0,1] \to [0,1]$ such that $\sigma$ and $\sigma^{-1}$ are Borel measurable and preserve the Lebesgue measure, such that $g(x,y)=f(\sigma(x),\sigma(y))$. The set of equivalence classes is denoted by 
$\widetilde{\mathcal{W}}$, while $\widetilde{f}$ is the equivalence class containing $f\in \mathcal{W}$.

In reviewing the known results in the thermodynamic limit, we follow here \cite{chatterjee2013estimating} 
(for a more general overview of large deviations for random graph see \cite{chatterjee2017large}).
The thermodynamic limit of  the pressure of the 
edge-triangle model  is given by \cite[Theorem 3.1]{chatterjee2013estimating} 
\be
\label{chap1:eq_var_problem_tr_edges}
\psi\left(\frac{h_{p}}{2},\frac{\alpha}{6}\right)
=\frac{1}{2}\sup_{\tilde{f}\in\mathcal{\widetilde{W}}} \left[\alpha\frac{t(\tilde{f})}{3} - I_{p}(\tilde{f}) \right] -\frac{\ln(1-p)}{2},
\ee
where $t(\tilde{f})$, the {\em density of triangles} in $\widetilde{f}$, is
\be
\label{chap1:def:density:triangles}
t(\tilde{f})= \int_{0}^{1}\int_{0}^{1}\int_{0}^{1} f(x,y)f(y,z)f(z,x)\,dx\,dy\,dz,
\ee 
and the entropic term 
is
\begin{equation}
I_{p}(\tilde{f})= \int_{0}^{1}\int_{0}^{1} I_p(f(x,y)) dxdy.
\end{equation}
Here $f$ is a representative element of the equivalence class $\tilde f$
and $I_{p}(u)$ for $u\in [0,1]$ denotes the {\em Bernoulli relative entropy}
\begin{equation}
\label{chap1:eq_entropia_bernoulli}
I_{p}(u):= u\ln\frac{u}{p} + (1-u)\ln\frac{1-u}{1-p}.
\end{equation}
Then, from  \eqref{eq:funzg-pressure} and \eqref{chap1:eq_var_problem_tr_edges} we get 
\begin{equation}
\label{cgf_variational_prob}
\mu_{p}(\alpha):=\lim_{n\to\infty}\mu_{n,p}(\alpha)= \sup_{\tilde{f}\in\mathcal{\widetilde{W}}} H(\tilde{f}),
\end{equation}
with 
\be\label{cgf_variational_form}
H(\tilde{f})=\alpha\frac{t(\tilde{f})}{3} - I_{p}(\tilde{f}). 
\ee
When the set of maximizers only consists of constant functions we say that we are in the \textit{replica symmetric} phase. 
Conversely, if the elements of the maximizing set are non-constant functions, then we say that we are in the \textit{replica symmetry breaking} phase. 

The infinite-dimensional variational problem \eqref{cgf_variational_prob}, involving the non-linear functional $H$, 
has been analytically solved only in a region of the parameter values.
In particular \cite[Theorem 6.2]{chatterjee2013estimating} proves that for all $0<p<1$ the system is in the replica symmetric phase for $\alpha>-2$.
The variational problem is then reduced to a scalar one,  see \cite[Theorem 4.1]{chatterjee2013estimating},
i.e. for $\alpha>-2$ we have
\begin{equation}
\label{chap1:eq:probl_scalare}
\sup_{0\leq\,u\leq\,1}\left[\alpha\frac{u^{3}}{3} -I_{p}(u) \right]=\alpha\frac{(u^{*}(\alpha))^3}{3} -I_{p}(u^{*}(\alpha)) =: \mu^{RS}_p(\alpha),
\end{equation}
where $u^{*}(\alpha)$ is the optimizer 
that solves the fixed-point equation:
\begin{equation}
\label{eq_punto_fisso}
\frac{e^{\alpha\,u^{2} +h_{p}}}{e^{\alpha\,u^{2} +h_{p}}+1}= u, \quad u\in [0,1].
\end{equation}
From this, one infers that for \cla{$\alpha >-2$} a graph sampled from the edge-triangle model in the limit $n\to\infty$ 
will look like an \ER random graph with parameter $u^{*}(\alpha)$,
i.e. edges are independent from each others and present with a probability $u^{*}(\alpha)$  (we refer to \cite{bhamidi2008mixing,chatterjee2013estimating} for the precise statement).

As for the region $\cla{-\infty < \alpha \le -2}$ the solution of the variational problem \eqref{cgf_variational_form} is unknown.
The Euler-Lagrange equation giving the stationarity condition are given by the following equation, 
which is the generalization of \eqref{eq_punto_fisso}  to the case of non constant functions \cite[Theorem 6.1]{chatterjee2013estimating}: 
	\begin{equation}
	\label{chap1:eq_pto_fisso_variaz}
	f(x,y)= \frac{\exp{(\alpha\int_{0}^{1}f(x,z)f(z,y)dz +h_{p})}}{\exp{(\alpha\int_{0}^{1}f(x,z)f(z,y)dz +h_{p})}+1}\, . 
	\end{equation}
It has been proved that for $\alpha$ small enough, a graph sampled from the edge-triangle model no longer resembles an \ER graph.
In particular  \cite[Theorem 6.3]{chatterjee2013estimating} show that 
for $\alpha$ small enough  (and for any  value of $p$)
the functional $H(\tilde{f})$
is not maximized at any constant function. This result is based on the fact \cite[Theorem 7.1]{chatterjee2013estimating} that one  actually proves that in the limit
$\alpha\to-\infty$, the solution of the variational problem \eqref{cgf_variational_prob}  is provided by the so-called 
\textit{equi-bipartite graphon} defined as
	\begin{equation}
	\label{chap1:def:graphon_bip}
	g(x,y):=
	\begin{cases}
	0 & \text{if\quad} (x,y)\in \left[0,\frac{1}{2}\right]^2 \cup \left[\frac{1}{2},1\right]^2 \\
	p & \text{if\quad} (x,y)\in \left[0,\frac{1}{2}\right]\times \left[\frac{1}{2},1\right] \cup \left[\frac{1}{2},1\right]\times \left[0,\frac{1}{2}\right],
	\end{cases}
	\end{equation}
and in this limit one has
	\begin{equation}
	\label{chap1:largenegative_limit}
	\lim_{\alpha\to -\infty}\mu_{p}(\alpha)= H(g) = \frac{1}{2}\ln(1-p)=: \hat{\mu}_p\,.
	\end{equation}
For later use, we remark that using the constant graphon $f_{\alpha}$ defined by $f_\alpha(x,y) = u^*(\alpha)$ for all $x,y\in [0,1]^2$,
one obtains
\be
\lim_{\alpha\to -\infty}\mu^{RS}_{p}(\alpha)=\lim_{\alpha\to -\infty}H(f_{\alpha}) = \ln(1-p)=: \hat{\mu}^{RS}_p\,.
\ee
\cla{Thus, in the limit $\alpha\to-\infty$ the replica symmetric solution is wrong by a factor 2.}
\subsection{A conjecture}
The graphon \eqref{chap1:def:graphon_bip} is the infinite volume correspondent of the {\em equi-bipartite graph}. In the latter the $\n$ vertices are partitioned into two disjoint sets of equal size with no edges connecting two vertices belonging to the same set. Clearly, in such a graph triangles do not exist (more generally bipartite graphs do not contain odd cycles \cite[Theorem 4]{bollobas2013modern}).
Analogously, the density of triangles in the equi-bipartite graphon \eqref{chap1:def:graphon_bip} vanishes since $t({g})=0$. Thus,
as expected, the edge-triangle model in the limit $\alpha\to-\infty$ is free of triangles.  
%
%

Guided by the results of our numerical analysis that we illustrate below, we conjecture that for any fixed $0<p<1$,  there exists a unique finite and negative value $\alpha_c(p)$ 
such that, crossing $\alpha_{c}(p)$ from above, the optimizer of  $H$ \cla{defined in \eqref{cgf_variational_form}} switches from the constant function $f_{\alpha}$, identically equal to the solution of the fixed-point equation \eqref{eq_punto_fisso}, 
to the graphon:
\begin{equation}\label{eq:graphon_bip_alpha}
g_{\alpha}(x,y):=
\begin{cases}
p_{1}{(\alpha)} & \text{if\quad} (x,y)\in \left[0,\frac{1}{2}\right]^2 \cup \left[\frac{1}{2},1\right]^2 \\
p_{2}{(\alpha)} & \text{if\quad} (x,y)\in \left[0,\frac{1}{2}\right]\times \left[\frac{1}{2},1\right] \cup \left[\frac{1}{2},1\right]\times \left[0,\frac{1}{2}\right],
\end{cases}
\end{equation}
where  $p_{1}{(\alpha)} $ and $p_{1}{(\alpha)}$ are functions taking value in $(0,1)$ that satisfy the following conditions:
\begin{align}
\lim_{\alpha\to-\infty}p_{1}{(\alpha)}=0,\label{eq:limpi}\\
\lim_{\alpha\to-\infty}p_{2}{(\alpha)}=p.\label{eq:limpb}
\end{align}
We observe that $\lim_{\alpha\to-\infty} g_{\alpha} = g$, with $g$ the graphon describing the equi-bipartite graph defined in \eqref{chap1:def:graphon_bip}. 
The rationale behind our conjecture is that the structure of  \eqref{eq:graphon_bip_alpha}  represents the  simplest geometry that may emerge from the breaking of the homogeneous graphon. 
\cla{For the resemblance of the structure of the overlap matrix in the 1RSB solution of 
spin-glasses}, we call this 
graphon the {\em 1-step replica symmetry breaking solution}. 

\section{\cla{Direct measurements for finite graphs}}\label{sectCloningRG}
In this section we 
\cla{compute numerically 
the cumulant generating function of the number of triangles
of a finite graph of size $n$. This expectation if substantially
affected by events that, although rare, give a large contribution
to the average defining the generating function.}
A standard tool \cla{for estimating the probability of rare events in the \ER random graph
is the importance sampling technique, see for instance \cite{bhamidi2015importance}}.
Here we follow an approach based on population dynamics, called  ``cloning'' \cite{giardina2006direct,giardina2011simulating,hurtado2014thermodynamics,perez2019sampling,carollo2019entanglement,angeli2019limit,angeli2019rare}.
%
%
%
%
%
\cla{In this section we adapt the method to a purely geometric problem, by introducing a dynamics for the graph construction}. 

\subsection{Implementing cloning}

The cloning algorithm \cla{is} obtained by tilting a Monte Carlo dynamics that samples from a target distribution.
In our case we would like to compute expectations w.r.t. the law of the  \ER graph \eqref{eq:prob-ee}.  
We first describe the dynamical process generating the \ER graph and then we recall how the tilted dynamic 
arises in the cloning algorithm. 

In the \ER graph each edge is present, independently,  with probability  $p\in (0,1)$. To generate a graph of size $n$,  we consider the Markov chain  $\{X_t, t\in \N \}$, taking values on the set $\kX$ of adjacency matrices,  defined as follows. We label the $n$ vertices in an arbitrary order. We start by selecting the first two vertices and we connect them with probability $p$, thereby obtaining a graph of size two. Then we select the third vertex and try to connect it to the first two vertices independently with probability $p$, thus obtaining a graph of size three. This procedure is repeated until the graph of size $n$ is formed: each time a new vertex is selected, it  is connected independently with probability $p$ to each of  those already visited. We stipulate that the discrete time step of this process corresponds to the attempt of adding a {\em single} new edge. Since the evolution from the graph of size $i$ to the graph of size $i+1$  requires $i$ attempts,  then the evolution starting  form size two and  leading to a graph of size $n$ will require  ${\mathcal N}_n= \sum_{i=2}^n i= {n \choose 2} -1$  steps.




We now introduce the tilted dynamics.
Denoting by $P$ the transition matrix of the Markov chain described above, i.e. $P(x,y)=\pp(X_{t+1}=y| X_t=x)$ for $x,y \in {\kX}$,   
the cumulant generating function of triangles reads
\eqan{
	&\mu_{n,p}(\za)  
	= \frac{1}{{\mathcal N}_n +1} \ln\left(\sum_{x_{0},\dots,\,x_{\kN_n} \in \kX_n}\nu_0(x_{0})P(x_{0},x_{1})\dots\,P(x_{\kN_n-1},x_{\kN_n})
	\e^{\frac{\alpha}{n}\,T(x_{\kN_n})}\right) \label{eq:path_graph},
}
where $\nu_0$ is the initial distribution (i.e. Bern$(p)$). 
We rewrite the number of triangles for the graph of size $n$ as
\be
T(x_{\kN_n})= \sum_{t=0}^{\kN_n} \Delta T (x_{t}, x_{t+1} ),
\ee
 where $\Delta T (x_{t}, x_{t+1} ) = T( x_{t+1} ) -T(x_{t})$ is the increment of triangles between two consecutive steps.
 In this way we obtain:
\eqan{
	&\mu_{n,p}(\za) =  \frac{1}{\kN_n +1} \ln\left(\sum_{x_{0},\dots,\,x_{\kN_n} \in \kX_n }\nu_0(x_{0})P(x_{0},x_{1})\dots\,P(x_{\kN_n-1},x_{\kN_n})
	\e^{\frac{\alpha}{n}\,\Delta T(x_{0},x_{1})}\dots\,\e^{\frac{\alpha}{n}\,\Delta T(x_{\kN_n-1},x_{\kN_n})}\right).
}
The average in the previous equation can be also computed according to a different dynamics. Indeed, introducing the quantity
\be\label{eq:cloning-factor}
k_\za(x)=\sum_{y\in {\mathcal \kX_n}} P(x,y) \e^{\frac{\alpha}{n}\,\Delta T(x,y)},\quad x \in \kX_n,
\ee
and the stochastic matrix $P_{\alpha}$ with elements 
\be\label{eq:palfat}
P_\alpha (x,y):=  P(x,y) \e^{\frac{\alpha}{n}\,\Delta T(x,y)}\frac{1}{k_\za(x)},
\ee
we can write:
\be\label{eq:mut2t}
\mu_{n,p}(\za) =\frac{1}{\kN_n+1}\ln\left(\sum_{x_{0},\dots,\,x_{\mathcal{T}_n} \in {\mathcal \kX_n}}\nu(x_{0})P_\za(x_{0},x_{1})\dots\,P_\za(x_{{\kN_n}-1},x_{{\kN_n}})
k_\za(x_{0})\dots\, k_\za(x_{\kN_n-1})\right) .
\ee
As observed in \cite{giardina2011simulating} this representation of the average  suggests a population dynamics scheme that, starting from a bunch of $M$ initial individuals (clones) with distribution $\nu_0(\cdot)$, makes them evolve according to the transition kernel  $P_{\alpha}(\cdot, \cdot)$ and reproduce according to the
rate $k_{\alpha}(\cdot)$.
Denoting by $M_{\kN_n}$ the size of the corresponding population of clones at the final time $\kN_n$, we get:
\be
\label{formulaCl}
\mu_{n,p}(\za) = \frac{1}{\binom{n}{2}}\ln \left [ \frac{M_{\kN_n}}{M}\right ].
\ee
\noindent
From the computational point of view, to avoid explosion or extinction,
at any time step $t$, the  family of $M_t$ of clones is brought back to the original size $M$ by picking uniformly at random $M$ clones out of the $M_t$ available. Recording  the ratio $R_t=M_t/M_{t-1}$, the left hand side of the previous display can be written in a telescopic form:
\be
\label{telesc_sum_cloning}
\mu_{n,p}(\za) = \frac{1}{\binom{n}{2}} \ln \left [ \prod_{t=3}^{\kN_n}R_t\right ] =  \frac{1}{\binom{n}{2}}  \sum_{t=3}^{\kN_n} \ln R_t.
\ee
Summarizing, formula  \eqref{eq:mut2t} is implemented, starting from  a population of $M$ elements randomly chosen  according to $\nu_0$, by iterating  
the following steps:
%

\begin{itemize}

	\item [a)] 
		Evolve independently  each clone (i.e. graph) with the transition probability $P_\alpha$.
	\item [b)] Replicate each clone with reproduction rate $k_{\alpha}$ thus obtaining a population of size $M'$.
	\item[c)] 
	The  ratio $R=M'/M$ is recorded and then the total number of copies is brought back to $M$, uniformly choosing $M$ clones among the $M'$.
\end{itemize}
\col{For a fixed size $M$ of the family of clones, the computational complexity of this algorithm is  $O(n^3)$. Indeed, the workload in the algorithm is essentially concentrated on the computation of the number of triangles $T(x)$. Since the increment of $T(x)$ in going from  size $i$ 
to size $i+1$ requires $i^2$ operations (being the number of triangles in the smaller graph known from the previous step), the overall cost for evaluating  the total number of triangles in the final graph of size $n$ will be  $\sum_{i=2}^n i^2=O(n^3)$.
Of course, to reach a prescribed level of accuracy, we have to choose the size of the cloning family as an increasing function 
of $n$. This has been discussed  in \cite{angeli2019limit,angeli2019rare}.}

\medskip\noindent
The procedure is further illustrated with the example of size $n=3$ in the Appendix.

\subsection{Numerical results} 
We present here the results of the cloning algorithm for the cumulant generating function of the triangles. 
We fix a population of $M=7000$ clones and a value of $p=0.4$, and we vary the variable $\alpha$ and the graph size $n$. 
We denote by $\mu^{Cl}_{n}(\alpha)$ the cumulant generating function returned by the cloning algorithm for the graph of size $n$ with $p=0.4$. Similarly, we shorthand $\mu^{RS}(\alpha):=\mu^{RS}_{0.4}(\alpha)$
and we denote the asymptotic values  $\hat{\mu} := \hat{\mu}_{0.4}$ 
and $\hat{\mu}^{RS}:= \hat{\mu}_{0.4}^{RS}$.
The outcomes of our simulations show that: 
\begin{itemize}
\item 
for  large values of $\alpha$, the algorithm reproduces 
the replica symmetric solution (i.e. $\mu^{Cl}_{n} \to \mu^{RS}$ as $n$ increases);
\item
for  small values of $\alpha$, the algorithm provides a cumulant generating function
which is independent of $\alpha$ within statistical fluctuations. This ``constant''  function
approximates better and better the value of the exact solution at $\alpha=-\infty$  (i.e. $\mu^{Cl}_{n} \to \hat{\mu}$ as $n$ increases);
\end{itemize}
We observe that the intersection between the replica symmetric solution and the exact solution
at $\alpha=-\infty$ occurs at a value $\alpha \simeq -110$, thus a substantially small
value where the rate of reproduction of triangles in the cloning algorithm is very small.
Despite this ``extreme" situation, we see strong evidence that the replica symmetry solution
does not hold for small $\alpha$'s, and for values $\alpha \leq -110$ the solution 
furnished by the equi-bipartite graphon is a very good approximation for the curve
returned by the algorithm.

%

\begin{figure}[htpb!]
	\begin{center}
		\subfigure[\label{rsymm_cl1}]%
		{\includegraphics[scale= 0.29]{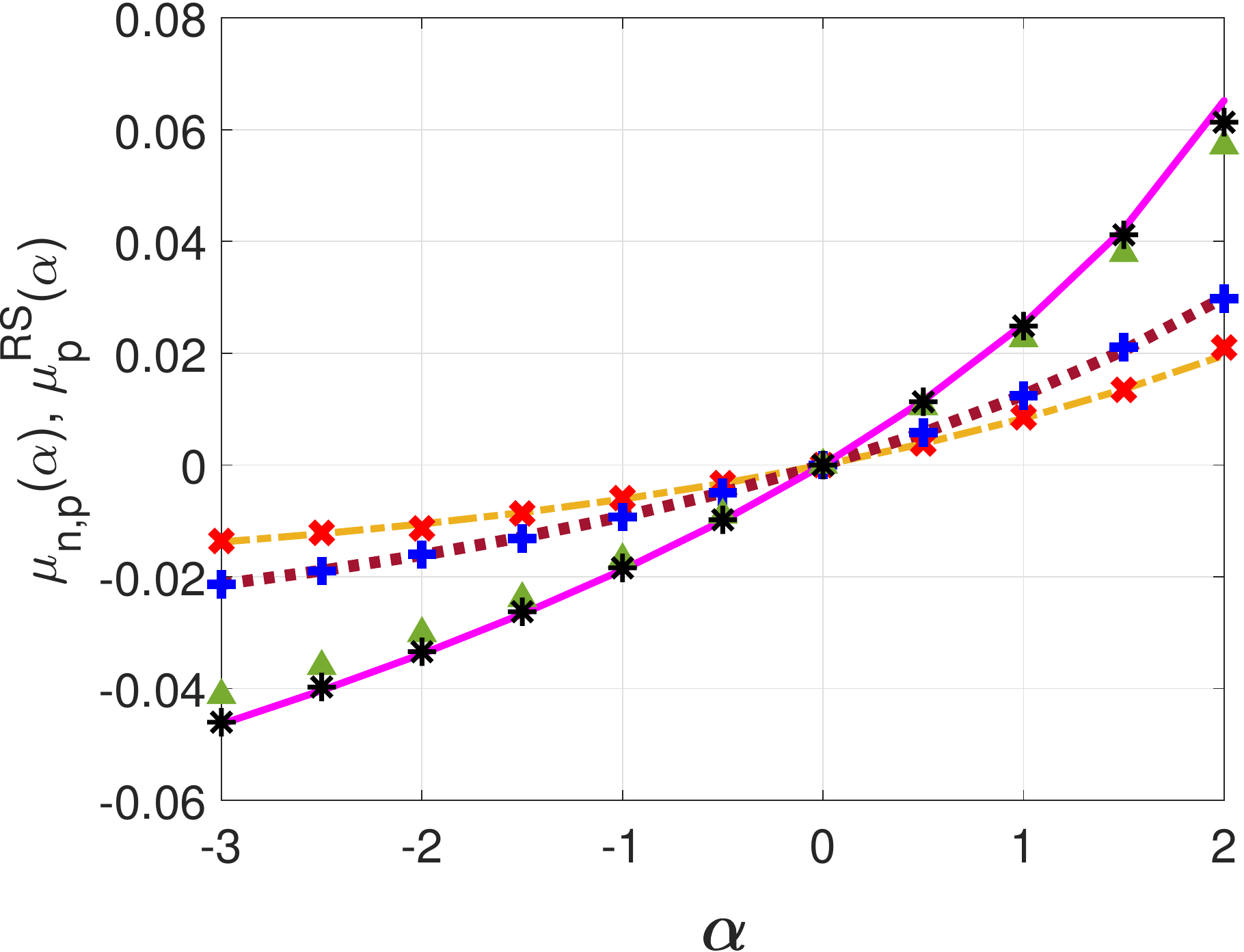}}
		 \subfigure[\label{rsymm_cl2}]
		{\includegraphics[scale= 0.29]{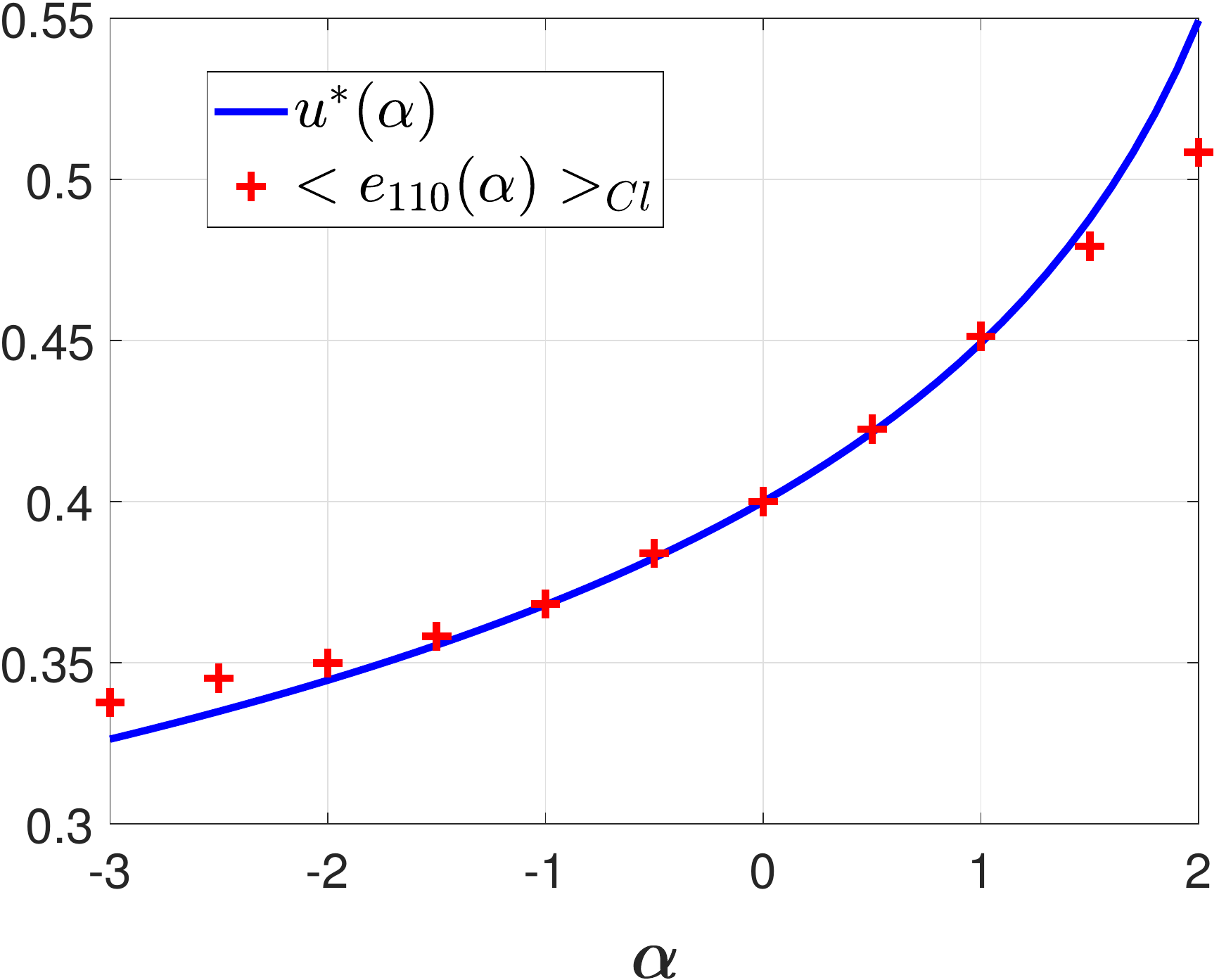}}
	      \subfigure[\label{rsymm_cl3}]
		{\includegraphics[scale= 0.29]{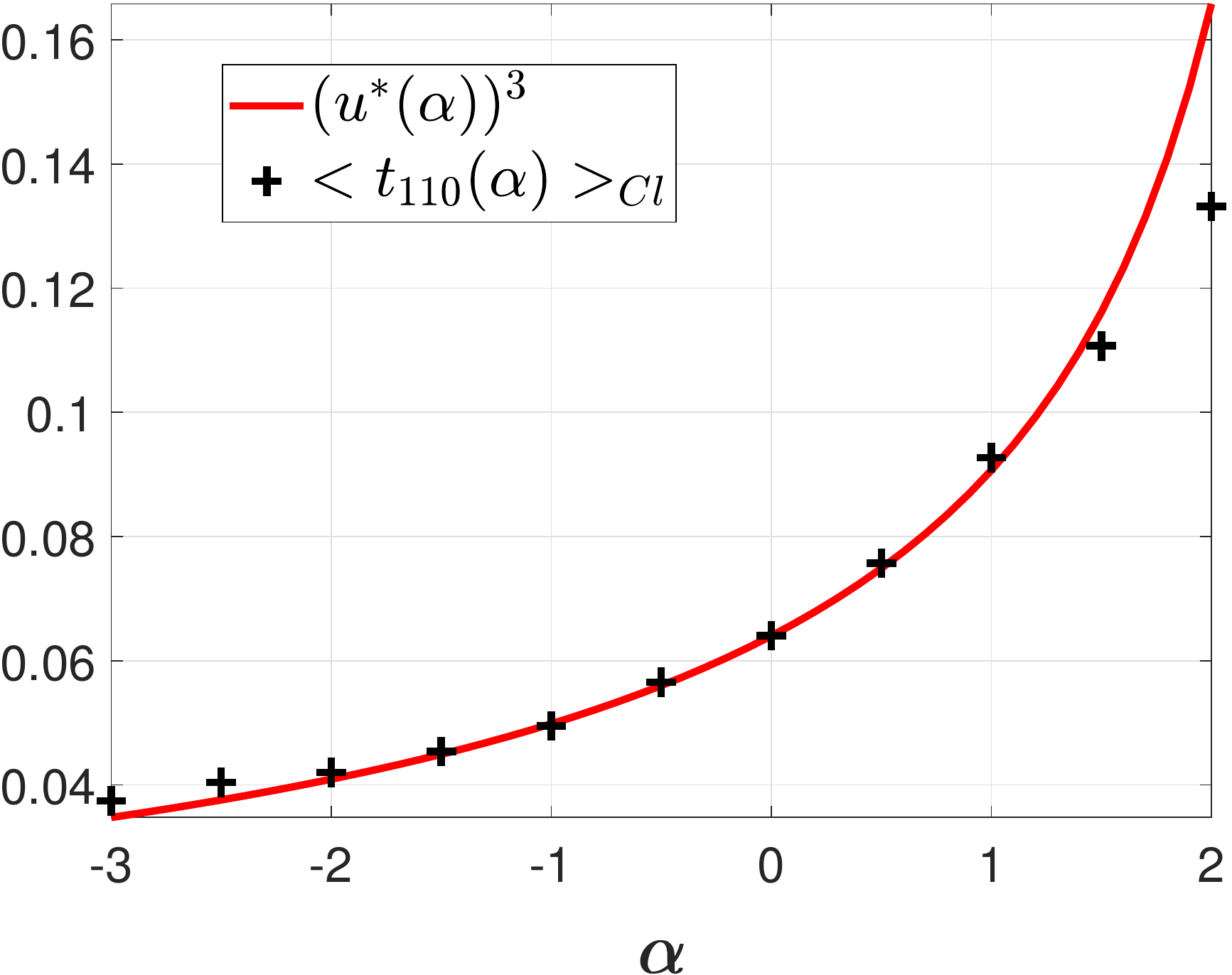}}
		\caption{Results of the cloning algorithm for $p=0.4$ for $\alpha\in[-3,2]$ (thus in the replica symmetric phase). Left panel: the picture shows the curves $\mu_{n,p}(\alpha)$ for $n=3$ (dashed-dotted yellow line), $n=4$  (dotted brown line) and the curve $\mu^{RS}_p(\alpha)$ (magenta continuous line), as well as the output of simulations $\mu^{Cl}_n(\alpha)$ represented  with dots: $n=3$ (${\bf x}$), $n=4$ (${\bf +}$), $n=20$ ($\blacktriangle$), $n=110$ (${\bf *}$). Central panel: normalized average number of edges for $n=110$ (+) in the clones ensemble, together with the solution $u^{*}(\alpha)$ of the fixed-point equation \eqref{eq_punto_fisso}. Right panel: normalized average number of triangles for $n=110$  (+) with $(u^{*}(\alpha))^3$ i.e. the expected density of triangles in  \ER model with parameter $u^{*}(\alpha)$. }\label{Cloning_a}.   
	\end{center}
\end{figure}

We discuss our results below.  Figure 1 explores the region around $\alpha=0$. 
Figure \ref{rsymm_cl1}
shows with dots the results of the cloning algorithm for $n=3$, $n=4$, $n=20$, $n=110$ and $\alpha\in[-3,2]$.
\begin{figure}[h!]
	\begin{center}
		\subfigure[\label{rb1}]%
		{\includegraphics[scale= 0.35]{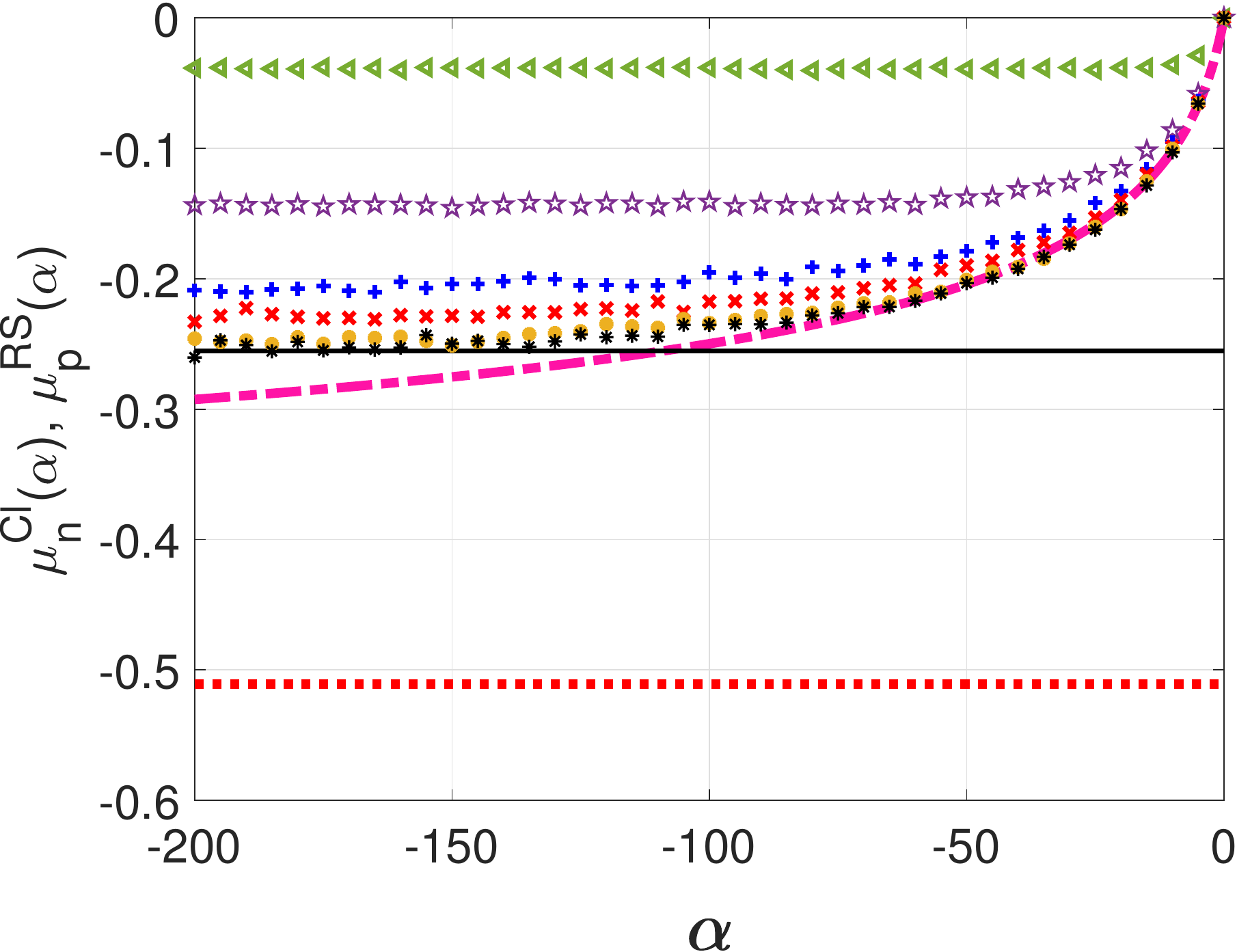}}
		\qquad \subfigure[\label{rb2}]
		{\includegraphics[scale= 0.35]{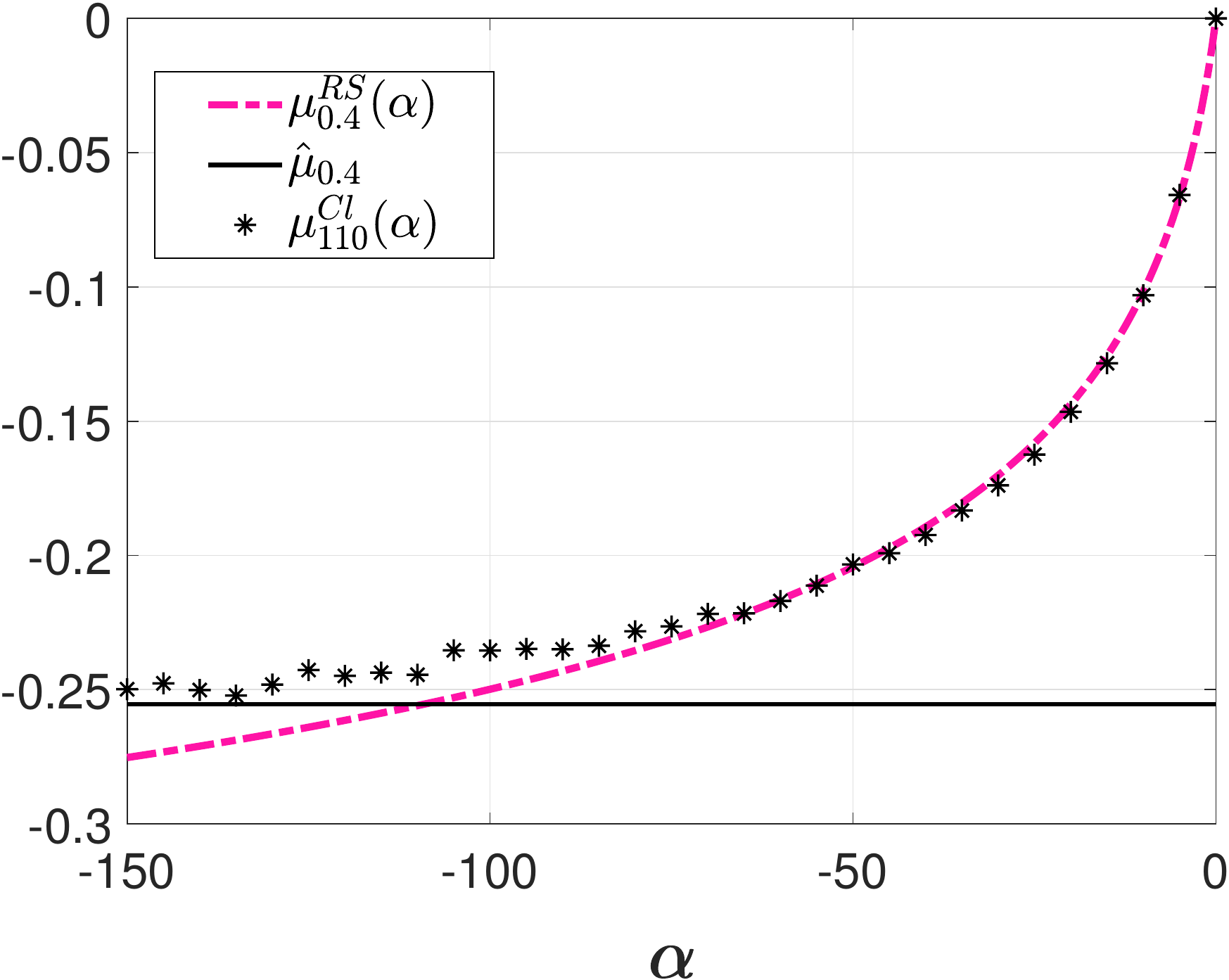}}\\
		\qquad \subfigure[\label{munAlfa1}]
		{\includegraphics[scale= 0.48]{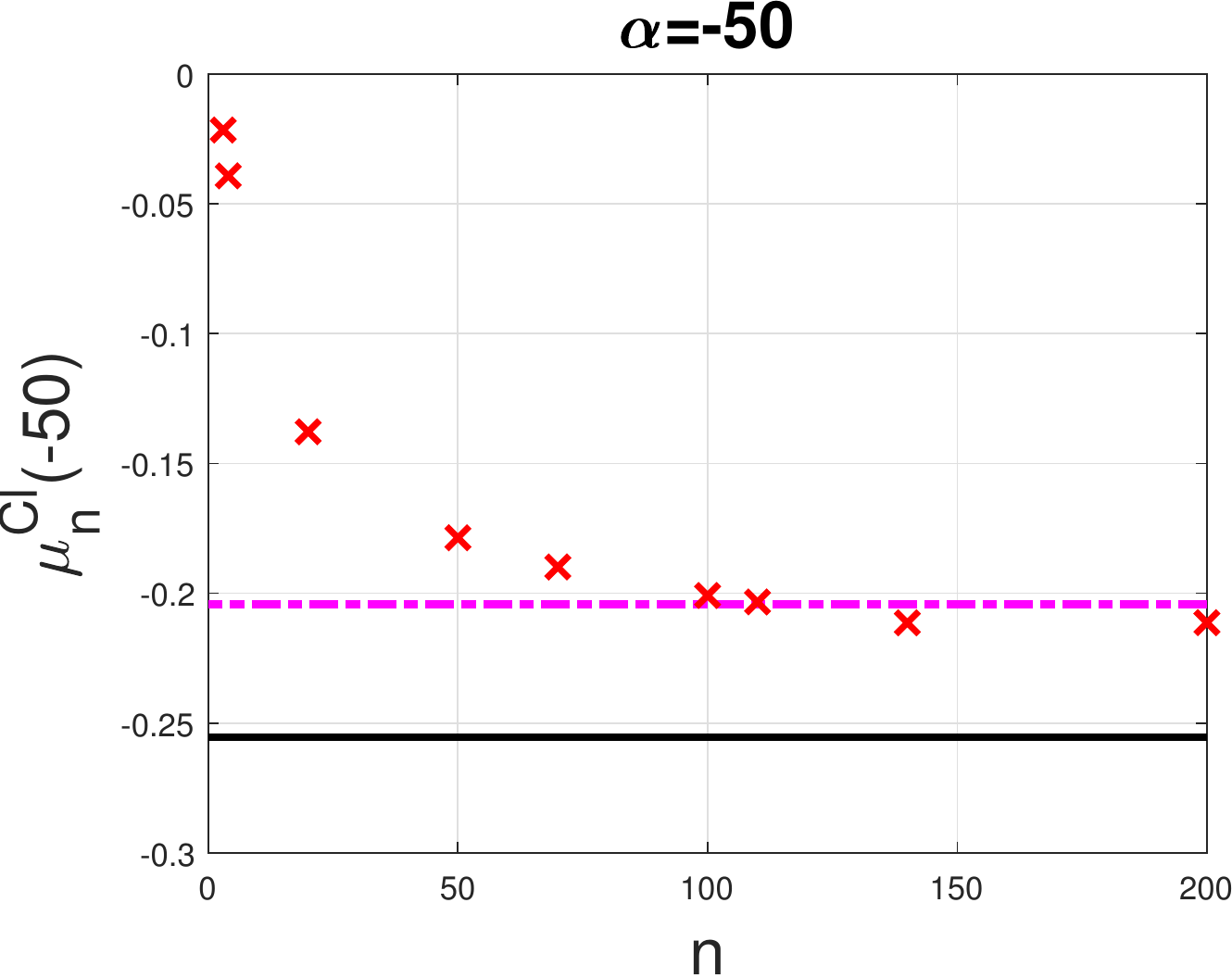}}
		\qquad \subfigure[\label{munAlfa2}]
		{\includegraphics[scale= 0.48]{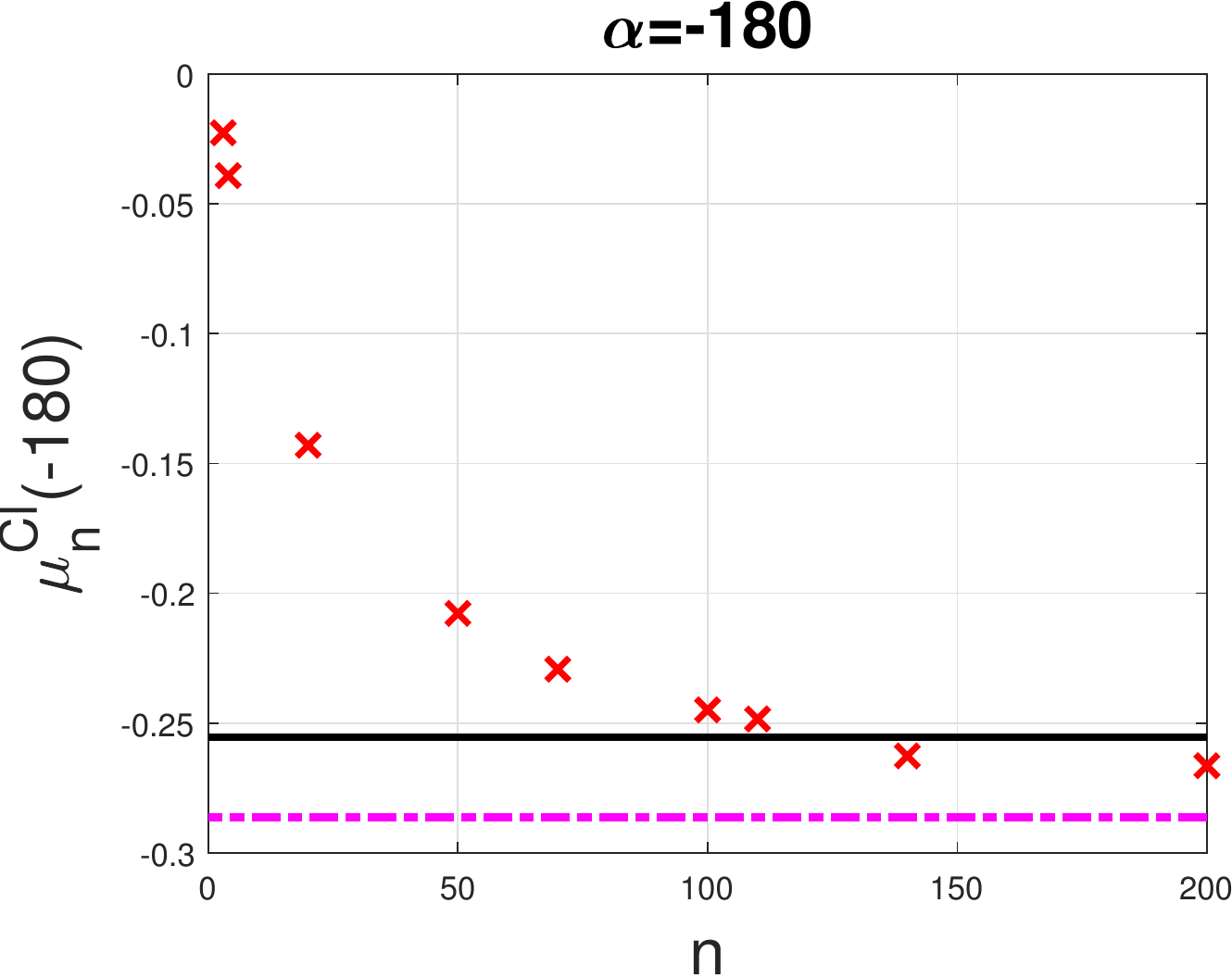}}
		\caption{Numerical results of the cloning algorithm for $p=0.4$.  Panel (a): the picture reports the numerical curves  $\mu^{Cl}_n(\alpha)$ with $n=4$ ($\blacktriangleleft$), $n=20$ ($\bigstar$), $n=50$ (+), $n=70$ (${\bf x}$), $n=100$ ($\bullet$), $n=110$ ($*$) together with the replica symmetric solution $\mu^{RS}_p(\alpha)$ (dashed-dotted magenta line), its asymptotic value  $\hat{\mu}^{RS}_p$  (dotted red horizontal line) and the true asymptotic  value  $\hat{\mu}_p$ (continuous black horizontal line). Panel (b): $\mu^{Cl}_n(\alpha)$ for $n=110$ ($*$) together with the replica symmetric solution $\mu^{RS}_p(\alpha)$ (dashed-dotted magenta line) and the asymptotic value $\hat{\mu}_p$ of $\mu_p(\alpha)$.
Panels (c) and (d): $\mu^{Cl}_n(\alpha)$ is plotted as a function of $n$ for two $\alpha$ values. 
In panel (c), where $\alpha=-50$, the points approach
the value $\mu^{RS}_{p}(-50)$  (dashed-dotted magenta horizontal line); in panel (d), where $\alpha=-180$, the points approach the value $\hat{\mu}_{p}$  (continuous black horizontal line).}
\label{Cloning_b}
	\end{center}
\end{figure}
\noindent
The picture also reports, with orange dashed-dotted  and maroon dotted lines, the exact curves $\mu_{n,0.4}$ for  $n=3$ and $n=4$, that can be computed explicitly.  The  curve $\mu^{RS}_p(\alpha)$, obtained by solving  numerically the scalar variational problem \eqref{chap1:eq:probl_scalare} characterizing the replica symmetric regime for $p=0.4$, is also displayed  (pink  continuous line).  We observe that for  $n=3$ and $n=4$ the cloning algorithm  perfectly reproduces the exact result and, as long as $n$ grows, the curves for increasing $n$'s settle on the curve  $\mu^{RS}(\alpha)$. 

A further check on the behavior of the algorithm is provided by Fig.\ref{rsymm_cl2}. It shows, as a function of $\alpha$,  the average density of edges  $\langle{e}_n(\alpha)\rangle_{Cl}$  in the population of clones  $x^{(i)}$,
that is 
$\langle{e}_n(\alpha)\rangle_{Cl} = \frac{1}{M }\sum_{i=1}^M  \frac{E(x^{(i)})}{{n \choose 2}}$.
This quantity  is an estimator of the  probability $p^\prime$ of  finding an edge connecting any two vertices of the graphs produced by the cloning. By effect of the tilting, $p^\prime$ is different from the original value  $p=0.4$ and is a function of $\alpha$, as expected. 
Since in replica symmetric regime the edge-triangle model  converges (in the limit $n\to \infty$) to an \ER model with parameter $u^{*}(\alpha)$, see Section \ref{sectKnownRes}, we have  reported in Fig.\ref{rsymm_cl2}  the optimizer  $u^{*}(\alpha)$.
The fair  agreement of the values   $\langle{e}_n(\alpha)\rangle_{Cl}$ on the curve of $u^{*}(\alpha)$, gives a further evidence that the cloning algorithm is well working in this
regime of $\alpha$ and reproduces the expected behavior of the edge-triangle model.  The same conclusion can be drawn form  Fig.\ref{rsymm_cl3} that shows the estimate  of the average density of triangles provided by cloning, i.e.   
$\langle{t}_n(\alpha)\rangle_{Cl} = \frac{1}{M }\sum_{i=1}^M  \frac{T(x^{(i)})}{{n \choose 3}}$  and the density of triangles in the symmetric replica regime, i.e. $(u^*(\alpha))^3$.

We now use the cloning method to explore the region with lower values of $\alpha$. We keep $p=0.4$ and
consider $\alpha\in[-200,0]$.
We  plot in Fig.\ref{Cloning_b} the results of the cloning method.
Panel (a) of Fig.\ref{Cloning_b} displays the cloning data as a function of $\alpha$ for several $n$ values, 
showing the converge towards a limiting profile as $n$ is increased. Besides the data points,
we also show the  replica symmetric cumulant generating function $\mu_{0.4}^{RS}(\alpha)$,
the asymptotic values $\hat{\mu}_{0.4}=-0.255$ and $\hat{\mu}_{0.4}^{RS}\simeq-0.510$. 
The replica symmetric solution $\mu_{0.4}^{RS}(\alpha)$ and the asymptote $\hat{\mu}_{0.4}$ (which we know is the
correct value of the cumulant generating function for $\alpha\to -\infty$) intersect around $\alpha=-110$. 
Panel (b) of Fig.\ref{Cloning_b}  exhibits the results for the largest graph size, i.e. $n=110$.
In panel (b) of Fig.\ref{Cloning_b} cloning data seem to be consistent with $\mu_{0.4}^{RS}(\alpha)$
for high enough values of $\alpha$, whereas they are flattening on top of  $\hat{\mu}_{0.4}$
for low $\alpha$ values. This is supported by the plots in panels (c) and (d), where we show
the $n$-dependence of the cloning data for two values of $\alpha$ (one below and one above
the intersection value $\alpha= -110$).  

\section{Numerical solution of the variational problem}\label{numsolxx}
The cloning algorithm implicitly solves the variational problem \eqref{cgf_variational_prob} by producing  a population of graphs that approximate the optimizing graphon. Unfortunately, it is hard to scrutinize the structure of the graphon from the adjacency matrix of the graphs, when their size is large. 
Thus, in order to study the graphon in the replica symmetry broken phase, we solve   \eqref{cgf_variational_prob}  via a numerical discretization. 
\subsection{Discretization}
The spatial discretization of  graphon  can be obtained considering the set of  $m\times\,m$ symmetric matrices $\{f_{i,j}\}_{i,j}$ with elements in $[0,1]$. 
However,  since the graphon that solves \eqref{cgf_variational_prob} does not take the values  0 and 1  \cite[Theorem 6.3]{chatterjee2013estimating}, we restrict our set of matrices to the following set:
\begin{equation}
\label{chap3:gamma}
\Gamma_{m,\varepsilon}:=\{f\in\mathbb{R}^{m\times\,m}: f_{ij}=f_{ji},\, f_{ij}\in[\varepsilon, 1-\varepsilon], \quad i,j=1,\ldots, m\},
\end{equation}
where $\varepsilon >0$ is a small parameter that bounds the functions  away from the singularity of the logarithm  in the discretization of $H(\tilde{f})$, that we define 
as follows:
\begin{equation}
\label{chap3:objective_f_dicr}
\mathcal{H}_{m}(f):= \frac{1}{m^2}\left[\frac{\alpha}{3m}\sum_{i,j,k=1}^{m}f_{ij}f_{jk}f_{ki} - \sum_{i,j=1}^{m}\left(f_{ij}\ln\frac{f_{ij}}{p} + (1-f_{ij})\ln\frac{1-f_{ij}}{1-p}\right)\right], \quad f\in\Gamma_{m,\varepsilon}.
\end{equation} 
We remark that, as it happens in the continuous setting,  $\mathcal{H}_{m}(f)$ enjoys a symmetry property. Indeed, given  $f,\, g \in \Gamma_{m,\varepsilon}$ we have $\mathcal{H}_{m}(g)=\mathcal{H}_{m}(f)$ if there exists a permutation  $\sigma$ over $\{1,\dots,m\}$ such that $g_{i,j}=f_{\sigma(i),\sigma(j)}$. In this case we call $g$ and $f$ \textit{equivalent}.

We solve numerically  the discretization  of \eqref{cgf_variational_prob}
\begin{equation}\label{probl_v_distcretizato}
\max_{f\in\Gamma_{m,\varepsilon}} \mathcal{H}_{m}(f)
\end{equation} 
by applying the \textit{Gradient Projection} (GP) method with both a constant and variable steplength, the latter chosen according to the Barzilai-Borwein rules \cite{barzilai1988two}. 
\subsection{Numerical results}
\begin{figure}[htpb!]
\begin{center}
\makebox[\linewidth]{
\begin{tabular}{cccc}
\subfigure[\label{fig3a}]
{\includegraphics[scale=0.23]{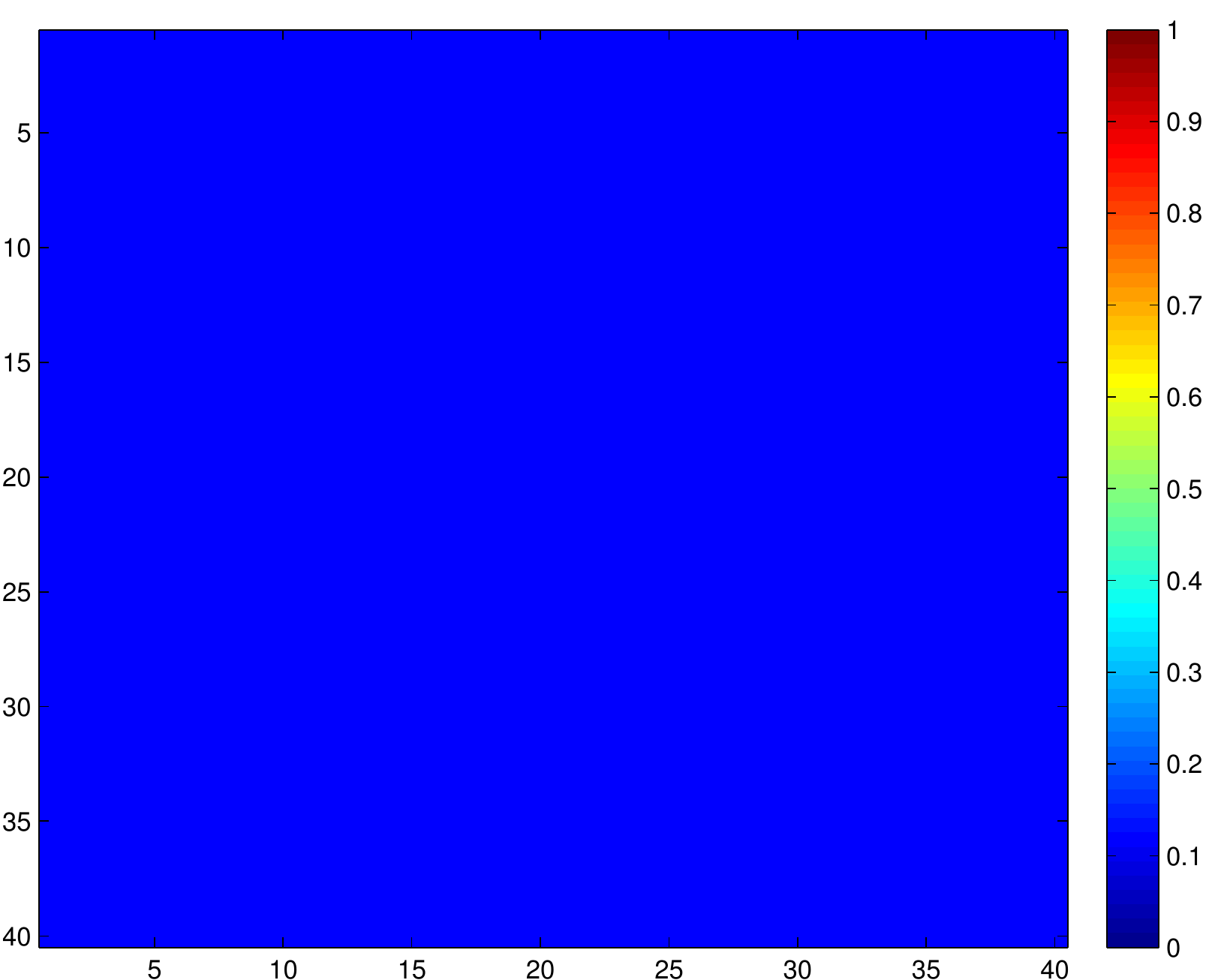}}
\subfigure[\label{fig3b}]
{\includegraphics[scale=0.23]{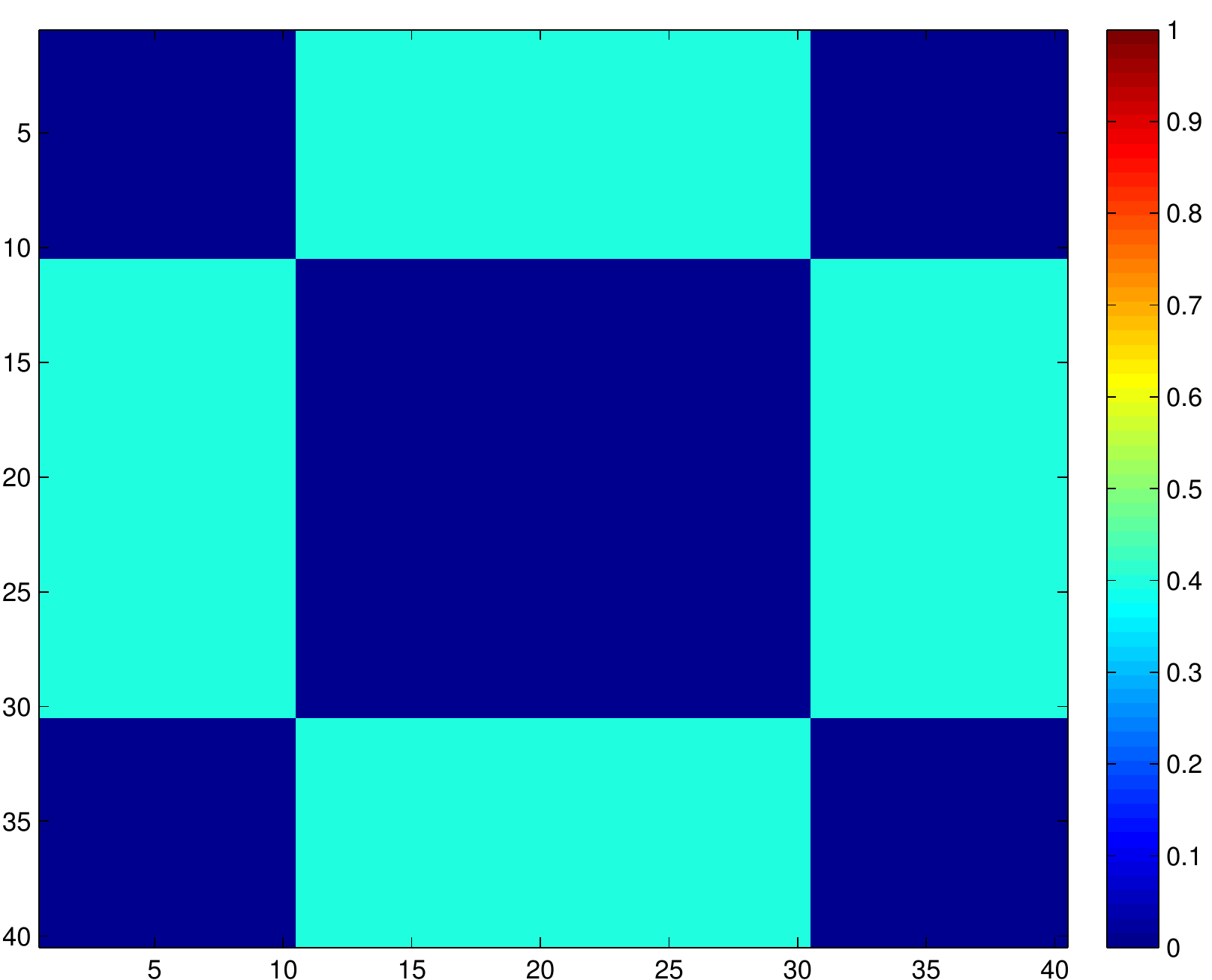}}
\subfigure[\label{fig3c}]
{\includegraphics[scale=0.23]{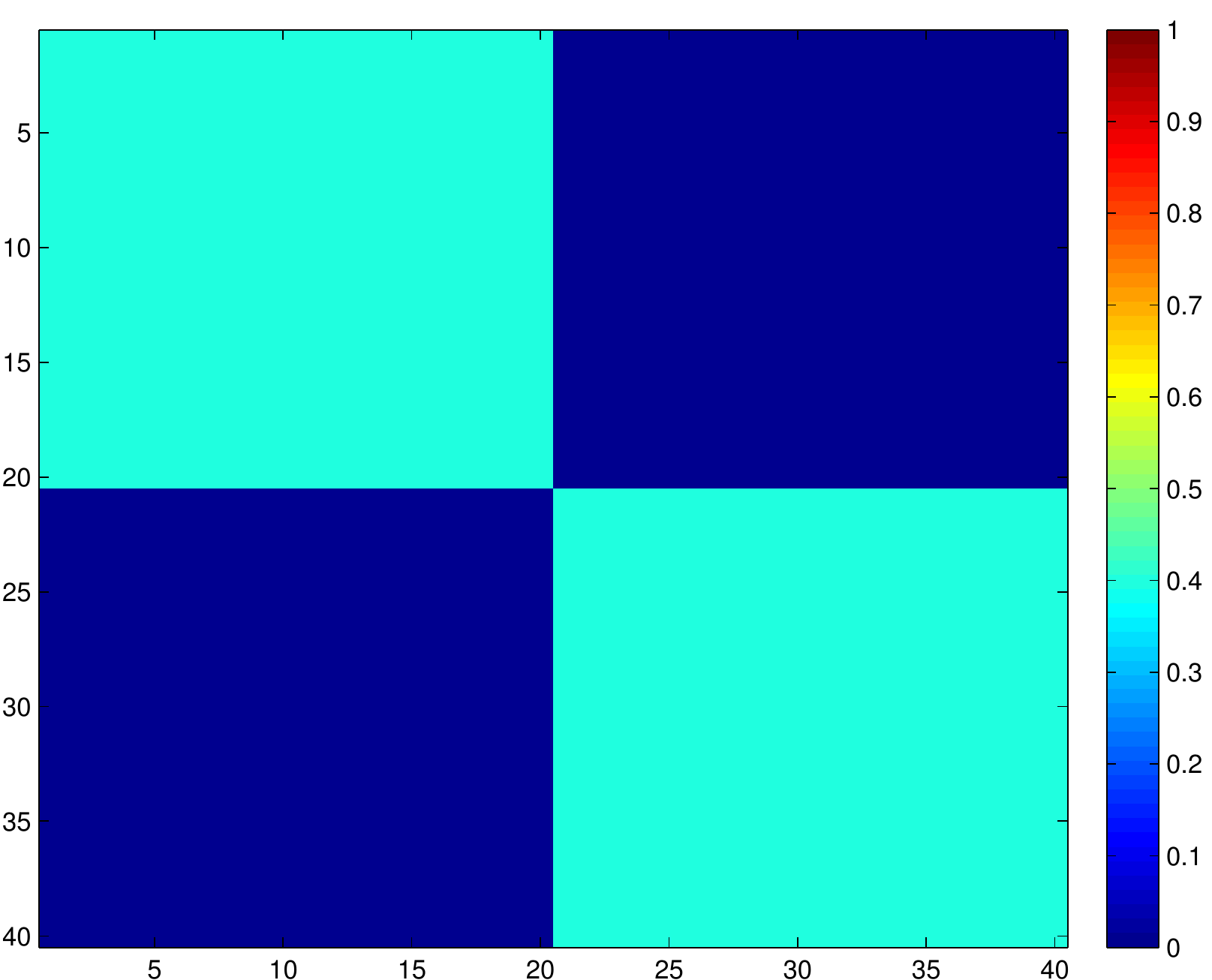}}
\subfigure[\label{fig3d}]
{\includegraphics[scale=0.23]{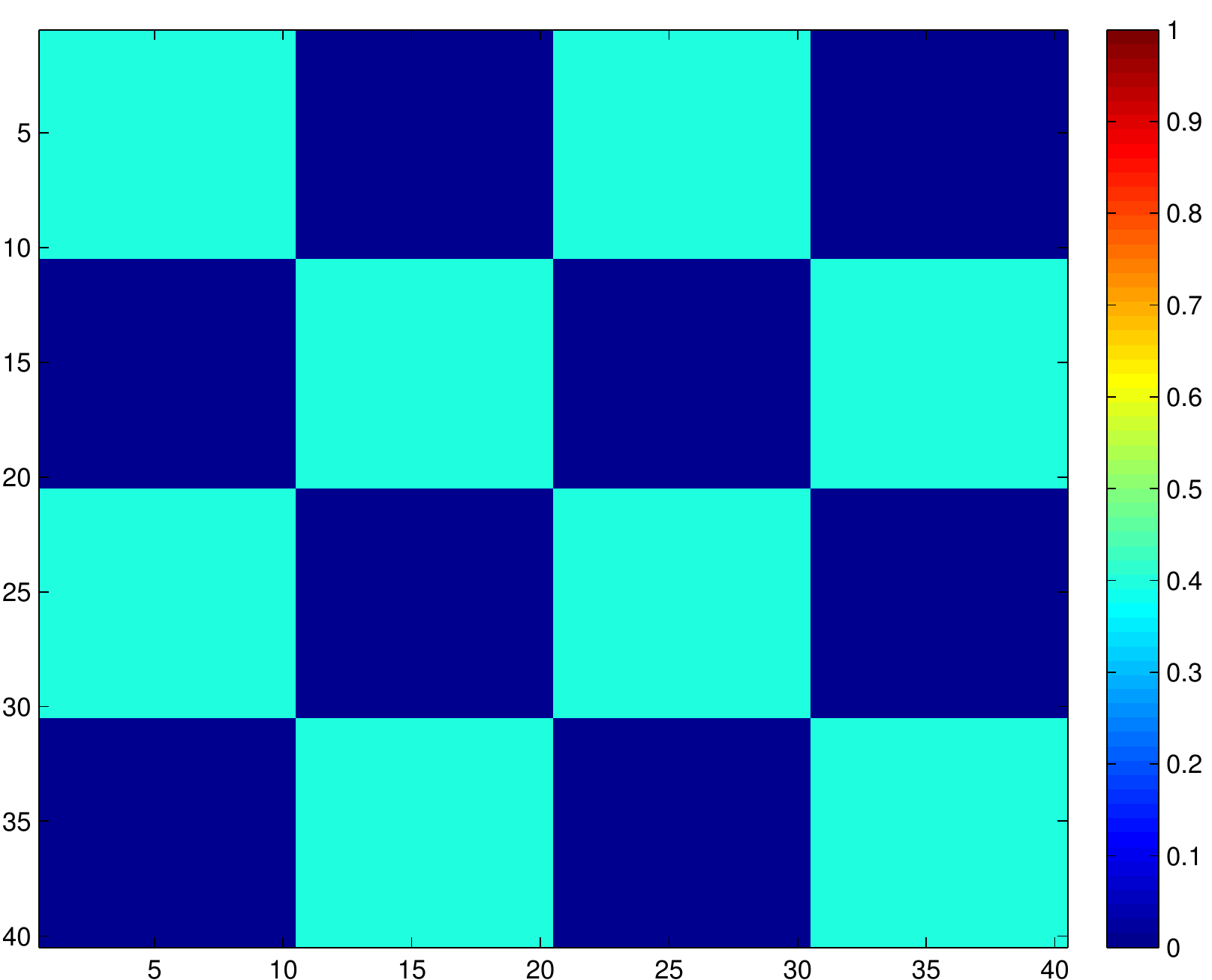}}
\end{tabular}}
\caption{The stationary points of \eqref{chap3:objective_f_dicr} returned by the GP method for $p=0.4$ and $\alpha=-110$. Different initial conditions, according to the steplength rules used, lead either  to the constant function (leftmost panel) and to a chessboard structure (the other panels). The value of the homogeneous solution on the leftmost panel is equal to $u^{*}(-110)$. The chessboard solutions are equivalent according to our definition.}\label{soluzioniGP}
\end{center}
\end{figure}

Being interested in the structure of the optimizer in the replica symmetry breaking region, we have solved \eqref{probl_v_distcretizato} for a set of $p$ values ranging from $0.2$ to $0.6$ and $\alpha$ below $-2$, varying it with unitary step. For each value of $p$ and $\alpha$ we have started the iterations of the GP method from a set of $12$ initial conditions, using a grid of size $m=40$ and setting $\varepsilon=10^{-4}$ (we tried other values of $m$ obtaining the same results).
Our main findings are:  
\begin{itemize}
\item[a)] the presence of only two different geometrical structures of maximizers: the constant ones and chessboard-like one
(see Fig.\ref{soluzioniGP}). The values of the constant maximizers turn out to be equal, within the numerical approximation, to the solution of the fixed-point equation \eqref{eq_punto_fisso};
\item[b)] there exists a critical value $\alpha_{c}(p)$ such that when $\alpha>\alpha_{c}(p)$ the optimizer of $\kH_{m}(f)$ is constant whereas when $\alpha<\alpha_{c}(p)$ it assumes a chessboard-like structure.   
\end{itemize}
We observe that there are different chessboard-like structures that can be reached starting from different initial conditions, as shown in  Fig.\ref{soluzioniGP}. 
All of them are equivalent to the discretization of the 1-RSB graphon, that by abuse of notation we also denote by $g_{\alpha}$:
\begin{equation}
\label{grafone_bip}
g_{\alpha}(i,j)=
\begin{cases}
p_{1} & \text{if\quad} (\frac{i}{m},\frac{j}{m})\in \left[0,\frac{1}{2}\right]^2 \cup \left[\frac{1}{2},1\right]^2, \\
p_{2} & \text{if\quad} (\frac{i}{m},\frac{j}{m})\in \left[0,\frac{1}{2}\right]\times \left[\frac{1}{2},1\right] \cup \left[\frac{1}{2},1\right]\times \left[1,\frac{1}{2}\right].
\end{cases}
\end{equation}



With $p$ fixed,  and varying $\alpha$, we sought the critical value $\alpha_{c}({p})$ by comparing the values of $\kH_m$ computed at the  
homogeneous solutions  $f_\alpha$ and at the chessboard-like ones  $g_\alpha$.   
We found that the values of $f_\alpha$ coincide, within our approximation and for all $\alpha$,  with $u^{*}(\alpha)$ (the solution of the replica symmetric fixed-point equation \eqref{eq_punto_fisso}), while  the values $p_{1}(\alpha)$ and $p_{2}(\alpha)$  occurring in $g_\alpha$ are close to $0$ and $p$, see Tab.1 for large negative $\alpha$ values. This implies that $g_\alpha$ is close to the equi-bipartite graphon  $g$ defined in \eqref{chap1:def:graphon_bip}. The evidence that the phase transition occurs  at the critical value $\alpha_c (p)$ is given by observing that  $\kH_m(f_\alpha)>\kH_m(g_\alpha)$ for  $\alpha >\alpha_c (p)$, while   $\kH_m(f_\alpha)<\kH_m(g_\alpha)$ for $\alpha <\alpha_c (p)$. An example is given in Tab.1 where, for the case $p=0.4$,  the change of the optimizer shows that the position of the critical value $\alpha_{c}({p})$ is between $-109$ and $-110$ (we recall that $\alpha$ varies with unitary step). 
\begin{table}[h!]
\begin{center}
\begin{tabular}{|c|c|c|c|c|c|c|}
$\alpha$ &$u^{*}(\alpha)$ & $f_\alpha$ & $g_\alpha$& $\kH_{40}(f_\alpha)$ & $\kH_{40}(g_\alpha)$\\
\hline
$-109$ & 0.120486 & 0.120486 & $p_{1}\approx\,0.000114,p_{2}\approx\,0.398829$  & -0.255336& -0.255356 \\
\hline
$-110$ &0.120082 & 0.120082 & $p_{1}\approx\,0.000100,p_{2}\approx\,0.399100$ & -0.255916 & -0.255361   
\end{tabular}
\end{center}
\caption{Numerical optimizers of \eqref{probl_v_distcretizato} for $p=0.4$ and $\alpha=-109,-110$. We report the solution $u^{*}(\alpha)$ to equation \eqref{eq_punto_fisso}, the constant solution $f_\alpha$ and the two values $p_{1}(\alpha)$ and $p_{2}(\alpha)$ taken by the chessboard solution $g_\alpha$. The last two columns show the values of  $\kH_{40}(f)$ that give evidence of the phase transition between $-109$ and $-110$.}\label{tab_riassunto_dati}
\end{table} 

\begin{figure}[h!]
\begin{center}
\makebox[\linewidth]{
\begin{tabular}{cc}
\subfigure[\label{fig4a}]%
{\includegraphics[scale = 0.3]{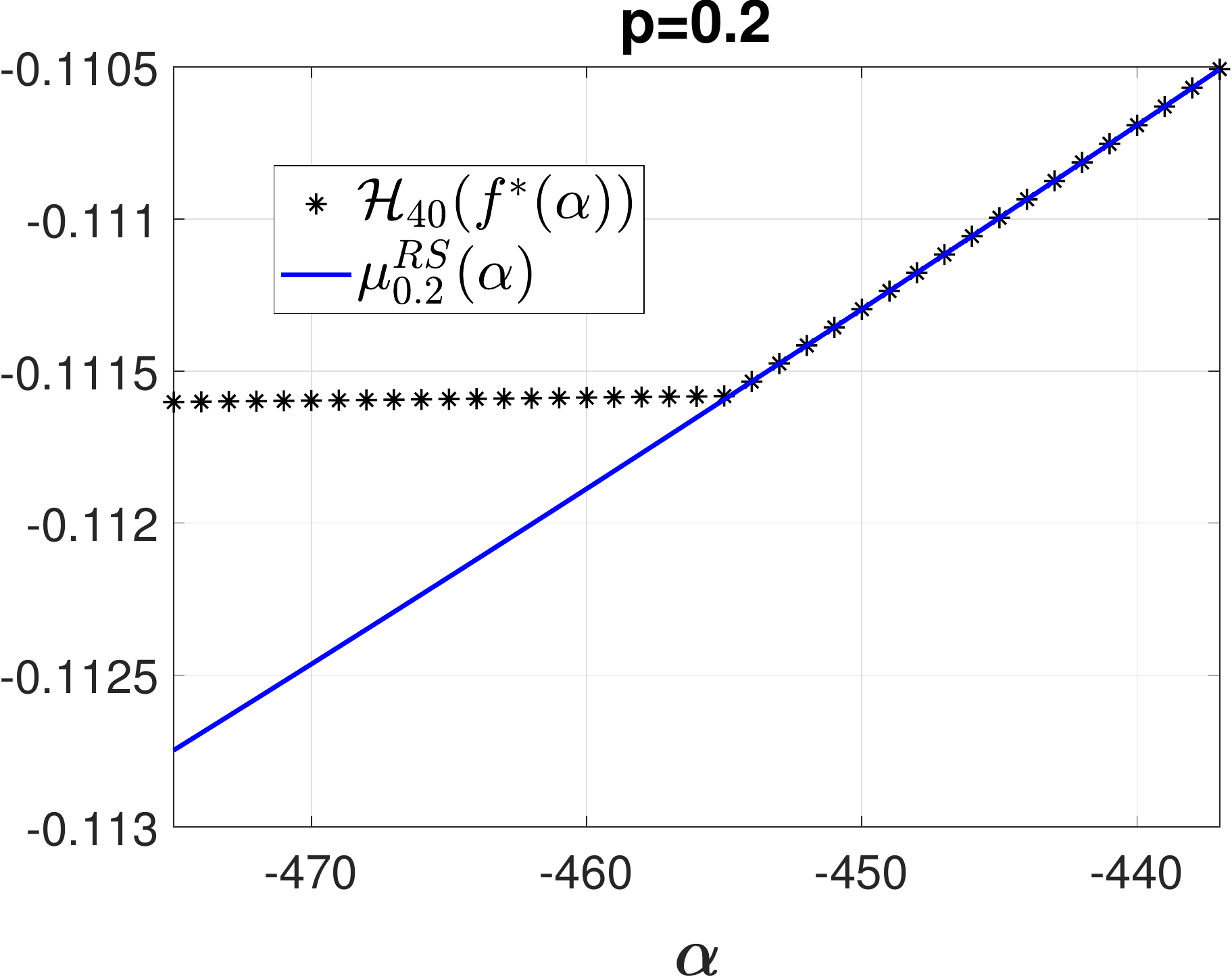}}
\subfigure[\label{fib4b}]
{\includegraphics[scale = 0.3]{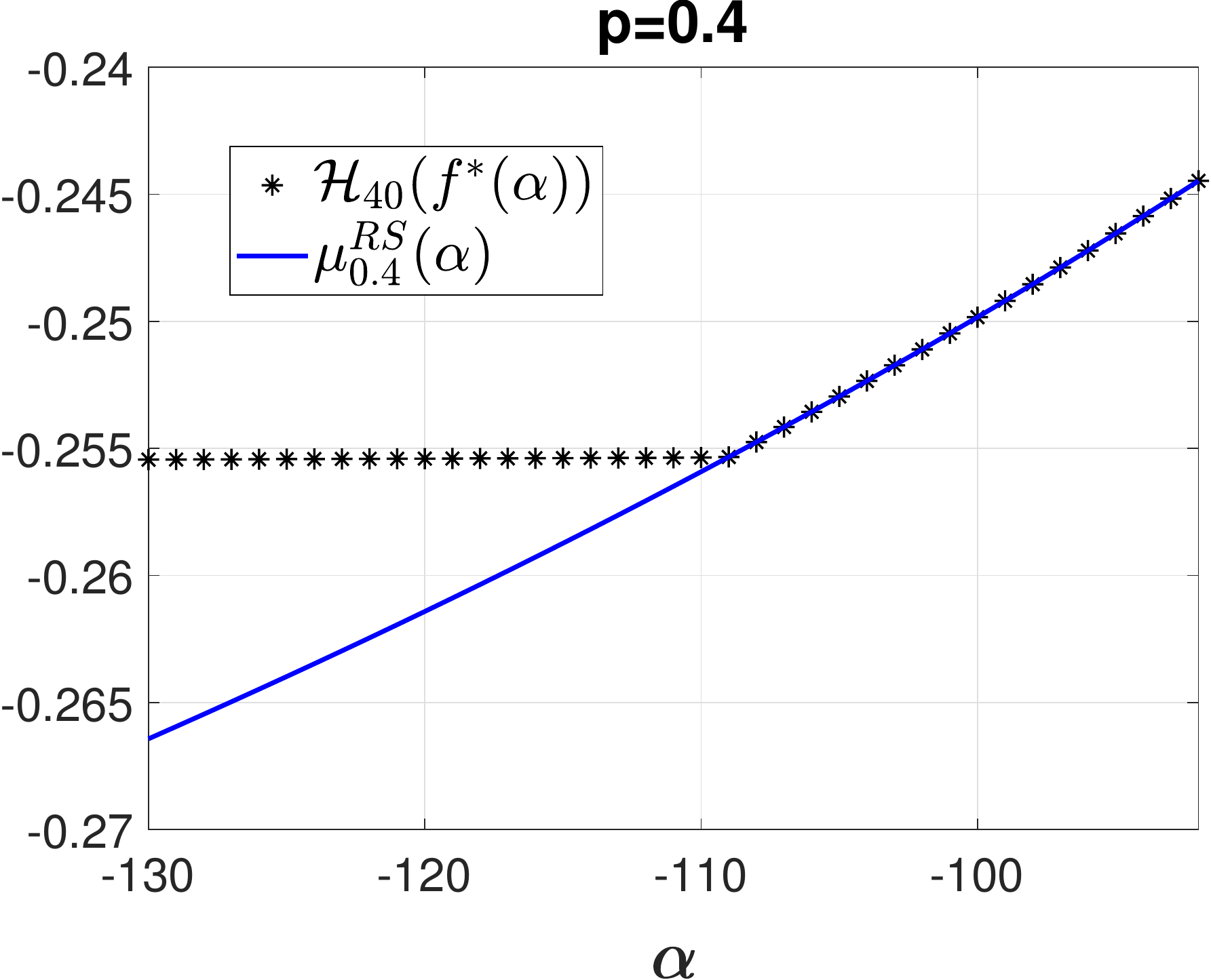}}
\subfigure[\label{fig4c}]
{\includegraphics[scale = 0.3]{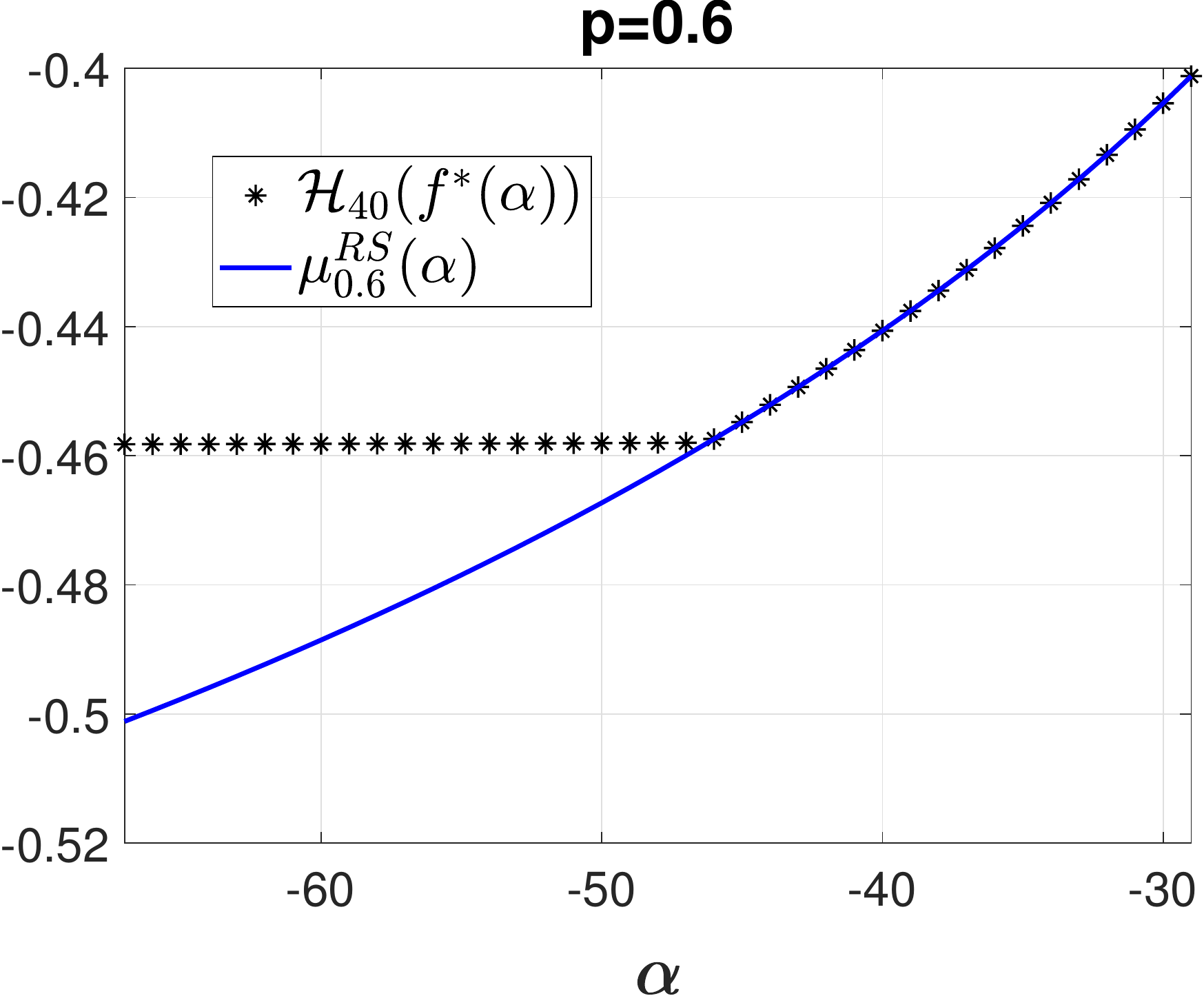}}
\end{tabular}}
\caption{Plot of $\mathcal{H}_{40}(f^{*}(\alpha))$ as a function of $\alpha$, where $f^{*}(\alpha)$ is the numerical solution of \eqref{probl_v_distcretizato}  (black, dotted line), together with the function ${\mu^{RS}_{p}(\alpha)}$ which solves problem \eqref{cgf_variational_prob} in the replica symmetric regime (continuous, blue line), {for $p=0.2,\, 0.4,\, 0.6$}.}\label{GP_method}
\end{center}
\end{figure}

Denoting by $h_\alpha$  the global optimizer found by the GP method, in  Fig.\ref{GP_method} we represent $\kH_{40}(h_\alpha)$ as a function of $\alpha$ for $p\in \{0.2,\, 0.4,\, 0.6\}$. 
In the replica symmetric region, i.e for $\alpha>\alpha_c({p})$, $\kH_{40}(h_\alpha)$ is expected  to approximate the solution $\mu^{RS}(\alpha)= \alpha\frac{{(u^{*}}(\alpha))^3}{3} -I_{p}(u^{*}(\alpha))$ .
The overlapping of the two curves above \col{$\alpha=-455$, $\alpha=-110$, and $\alpha=-47$} is shown in Fig.\ref{GP_method}. Below such thresholds, that we identify as  approximations of critical value $\alpha_{c}({p})$, the optimal solution $h_\alpha$ of $\kH_{40}$ switches from the constant one $f_\alpha$ (approximating the fixed-point equation \eqref{eq_punto_fisso}) to a chessboard function $g_\alpha$. 
The function   $\kH_{40}(h_\alpha)$  is nearly flat for $\alpha<\alpha_{c}(p)$ and very close to the asymptotic value  $H(g)=\frac 1 2 \ln(1-p)$ of $\mu_p(\alpha)$, see  \eqref{chap1:largenegative_limit}.

\section{Numerical solution of the restricted variational problem: the  1-step RSB optimizer}\label{sez1stepRB}

In this section, in order to study the transition between the replica symmetric region and replica symmetry breaking region, 
we return to the continuous variational problem \eqref{cgf_variational_prob}. 
In view of the numerical results of Section \ref{numsolxx}, here we {\em assume} that the breaking of the homogeneous graphon gives rise to a 
solution with the chessboard structure, see Fig.\ref{soluzioniGP},
\begin{equation}
\label{def:graphon_bip_ext}
g^{(a)}_{p_{1},p_{2}}(x,y)=
\begin{cases}
p_{1} & \text{if\quad} (x,y)\in \left[0,a\right]^2 \cup \left[a,1\right]^2 \\
p_{2} & \text{if\quad} (x,y)\in \left[0,a\right]\times \left[a,1\right] \cup \left[a,1\right]\times \left[0,a\right],
\end{cases}
\end{equation}
that we call the  {\em generalized 1-step replica symmetry breaking} solution. We introduce the set  
\begin{equation*}
{\cal R} =\{ g^{(a)}_{p_{1},p_{2}}(x,y) |\, a,\,  p_{1},\,  p_{2}\in [0,1] \} \subset {\cal W}.
\end{equation*}
This set contains the constant graphon  (obtained by setting $p_{1}=p_{2}$ for any $a$ or, equivalently, by setting either $a=0$ or $a=1$) and, for $a=\frac1 2$,  the graphon \eqref{eq:graphon_bip_alpha}, that we claim to be the solution of \eqref{cgf_variational_prob} for  $\za< \za_c(p)$.  Also, the limiting graphon  \eqref{chap1:def:graphon_bip} is contained in $\kR$.

We thus reduce the infinite dimensional problem  \eqref{cgf_variational_prob} to the finite dimensional one obtained by restricting to the set $\kR$:
 \begin{equation}\label{funzionale}
\sup_{g\in \kR }\left[ \frac{\alpha\,t(g)}{3}-I_{p}(g)\right] \equiv \sup_{p_{1},p_{2},a\in[0,1]}\left[ \frac{\alpha\,t(g^{(a)}_{p_{1},p_{2}})}{3}-I_{p}(g^{(a)}_{p_{1},p_{2}})\right]
=: \mu_{p}^{1RSB}(\alpha). 
\end{equation}

Being interested in the phase transition, we consider this problem for  $\alpha \leq -2$.
Using this approach, we aim at locating the critical value $\widetilde{\za}_c(p)$ denoting the transition that we claim to occur  in $\cal R$ between the  homogeneous and the 1-step replica breaking solution. Obviously we can state that $\widetilde{\za}_c(p) \le {\za}_c(p)$, but  we do not have any argument to assert  that the transition in $\cal R$ is the same  occurring in the whole space  $\cal W$ and, as a consequence,  that $\widetilde{\za}_c(p)$  coincides with ${\za}_c(p)$. However, some evidence for the equality of the two critical points will be obtained in this section.

{
The function to be maximized in \eqref{funzionale} can be written as follows:
\begin{align}
&\mathcal{F}_{\alpha}(p_{1},p_{2},a):=\alpha\frac{t(g^{(a)}_{p_{1},p_{2}})}{3} - I_{p}(g^{(a)}_{p_{1},p_{2}}) \notag\\
&= \frac{\alpha}{3}\{ p^{3}_{1}[a^{3}+(1-a)^3] + 3p^{2}_{2}p_{1}a(1-a)   \} - 2a(1-a)I_{p}(p_{2}) - [a^{2} + (1-a)^{2}]I_{p}(p_{1}), \notag
\hspace{-0.5cm}
\label{chap3:funzionale_graf_bip}
\end{align}
since the  density of triangles \eqref{chap1:def:density:triangles}  in the generalized 1-step replica symmetry breaking  graphons  \eqref{def:graphon_bip_ext} is 
\be\label{triangrsb}
t(g^{(a)}_{p_{1},p_{2}})=p^{3}_{1}[a^{3}+(1-a)^3] + 3p^{2}_{2}p_{1}a(1-a).
\ee
\noindent

 }
This function, that satisfies the symmetry $\mathcal{F}_{\alpha}(p_{1},p_{2},a)=\mathcal{F}_{\alpha}(p_{1},p_{2},1-a)$, can be defined by continuity up to the boundaries of  $[0,1]^3$ by setting $I_{p}(0)=I_{p}(1)=0$. Therefore  $\mathcal{F}_{\alpha}(p_{1},p_{2},a)$  attains its maximum on $[0,1]^3$. 
Due to the fact that the optimizer of \eqref{cgf_variational_prob} is bounded away from $0$ and $1$ \cite{chatterjee2013estimating}, we assume that the coordinates $(p_1,p_2)$ of the maximum point  lie in the interior of $[0,1]^2$ and thus satisfies  stationarity condition $\nabla\mathcal{F}_{\alpha}(p_{1},p_{2},a)=0$,  given by the following system of equuations:
\begin{equation}\label{chap3:sistema_Grad}
\begin{cases}
\alpha\left[p^{2}_{1}(3a(a-1)+1) + ap^{2}_{2}(1-a)\right] -(a^{2} +(1-a)^{2})\ln\left(\frac{p_{1}(1-p)}{p(1-p_{1})}\right)=0, &\quad (a)\\
2a(1-a)\left[p_{1}p_{2}\alpha -\ln\left(\frac{p_{2}(1-p)}{p(1-p_{2})} \right)\right]=0, &\quad (b)\\
(2a-1)\left[\alpha(p^{3}_{1} -p^{2}_{2}p_{1}) +2(I_{p}(p_{2}) -I_{p}(p_{1}))\right]=0, &\quad (c)
\end{cases} 
\end{equation}

\begin{figure}
	\begin{center}
		\makebox[\linewidth]{
			\begin{tabular}{ccc}
			 \subfigure[\label{fig5a}]
				{\includegraphics[scale= 0.3]{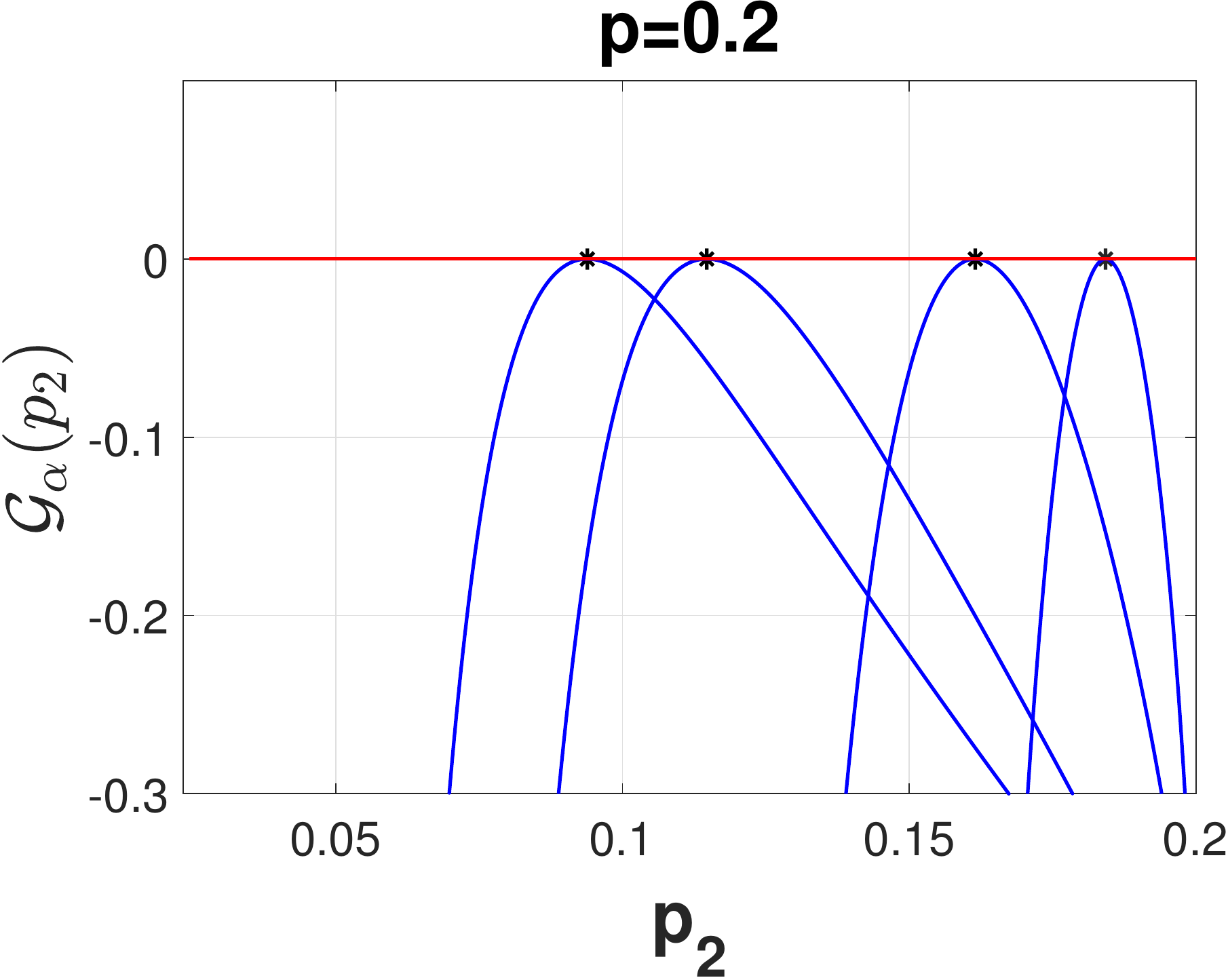}}
                  \subfigure[\label{fig5b}]
				{\includegraphics[scale=0.3]{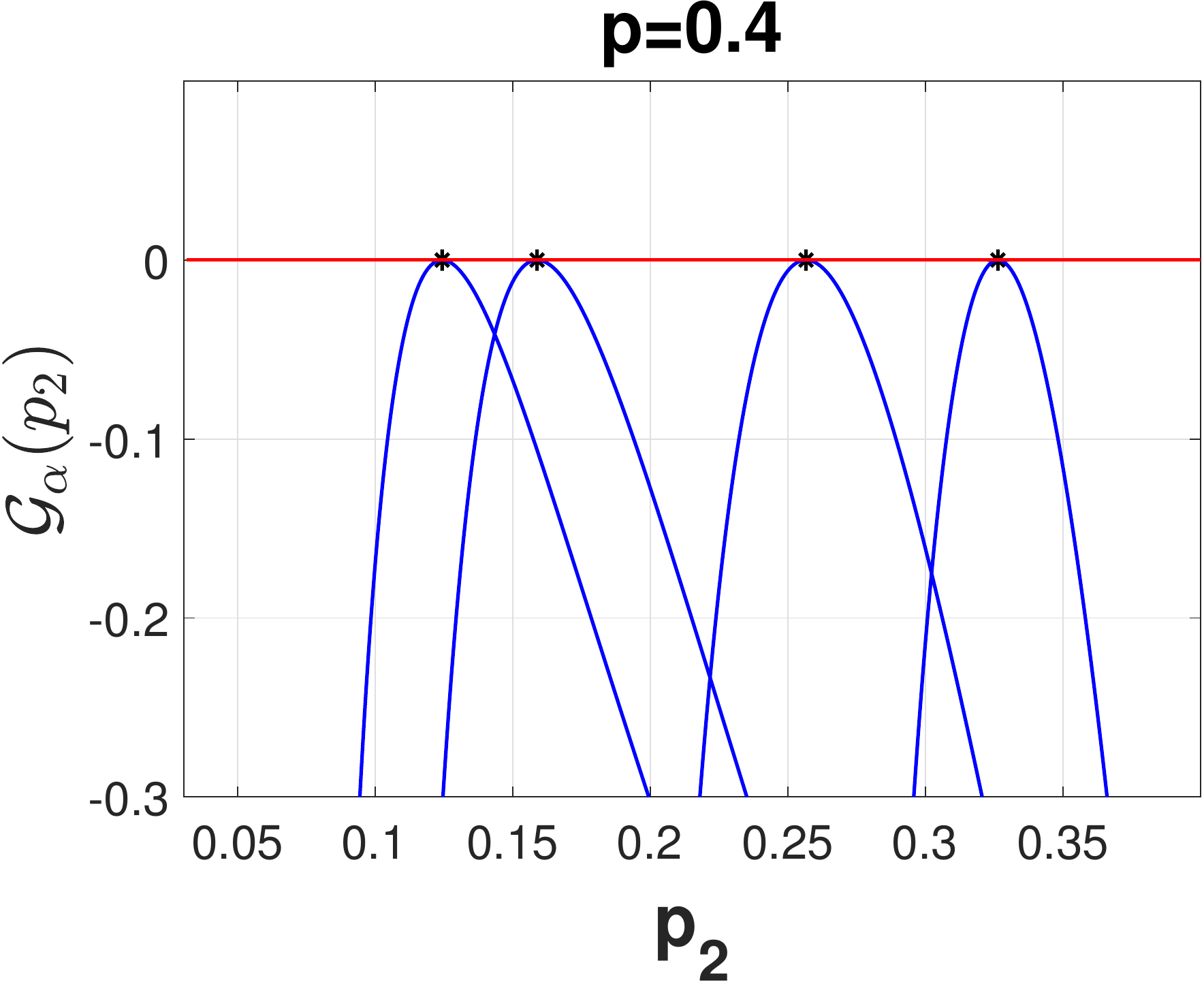}}
			\subfigure[\label{fig5c}]	
				{\includegraphics[scale=0.3]{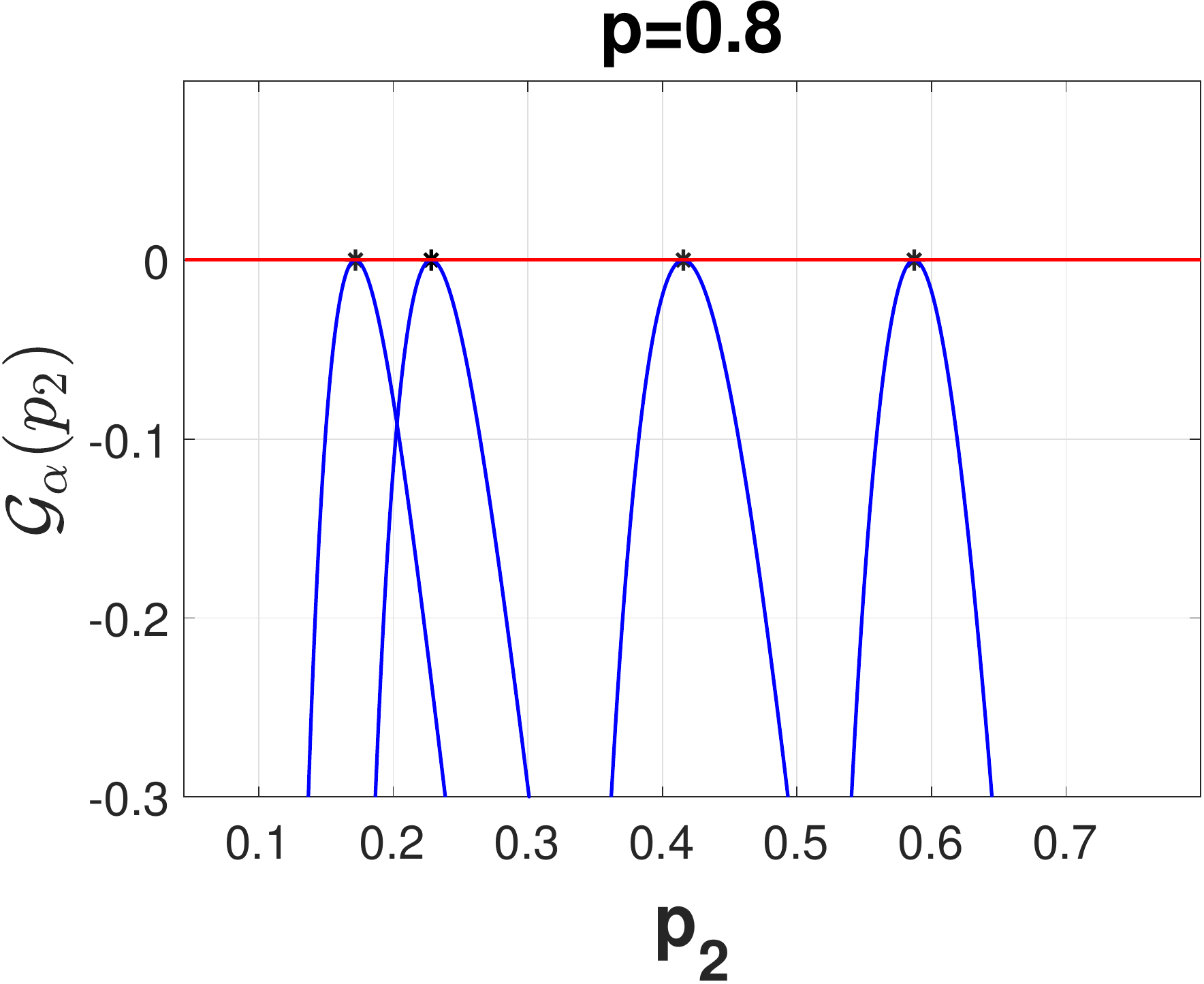}}
		\end{tabular}}
		\caption{Plot of the function $\kG_\alpha(p_2)$ defined in \eqref{equazg} for  $p=0.2,\, 0.4,\, 0.8$. In each panel the curves represent,  from right to left,  $\kG_\alpha(p_2)$ for  $\alpha=-3, -10, -50, -100$.  The black dots on  the horizontal  line $y=0$ represent $u^*(\alpha)$. These plots show that $u^*(\alpha)$ is the unique solution to equation  \eqref{equazg}.
		}\label{nuovafigura}
		\end{center}
\end{figure}
\begin{figure}[h!]
	\begin{center}
		\makebox[\linewidth]{
			\begin{tabular}{ccc}
			\subfigure[\label{fig6a}]
				{\includegraphics[scale= 0.3]{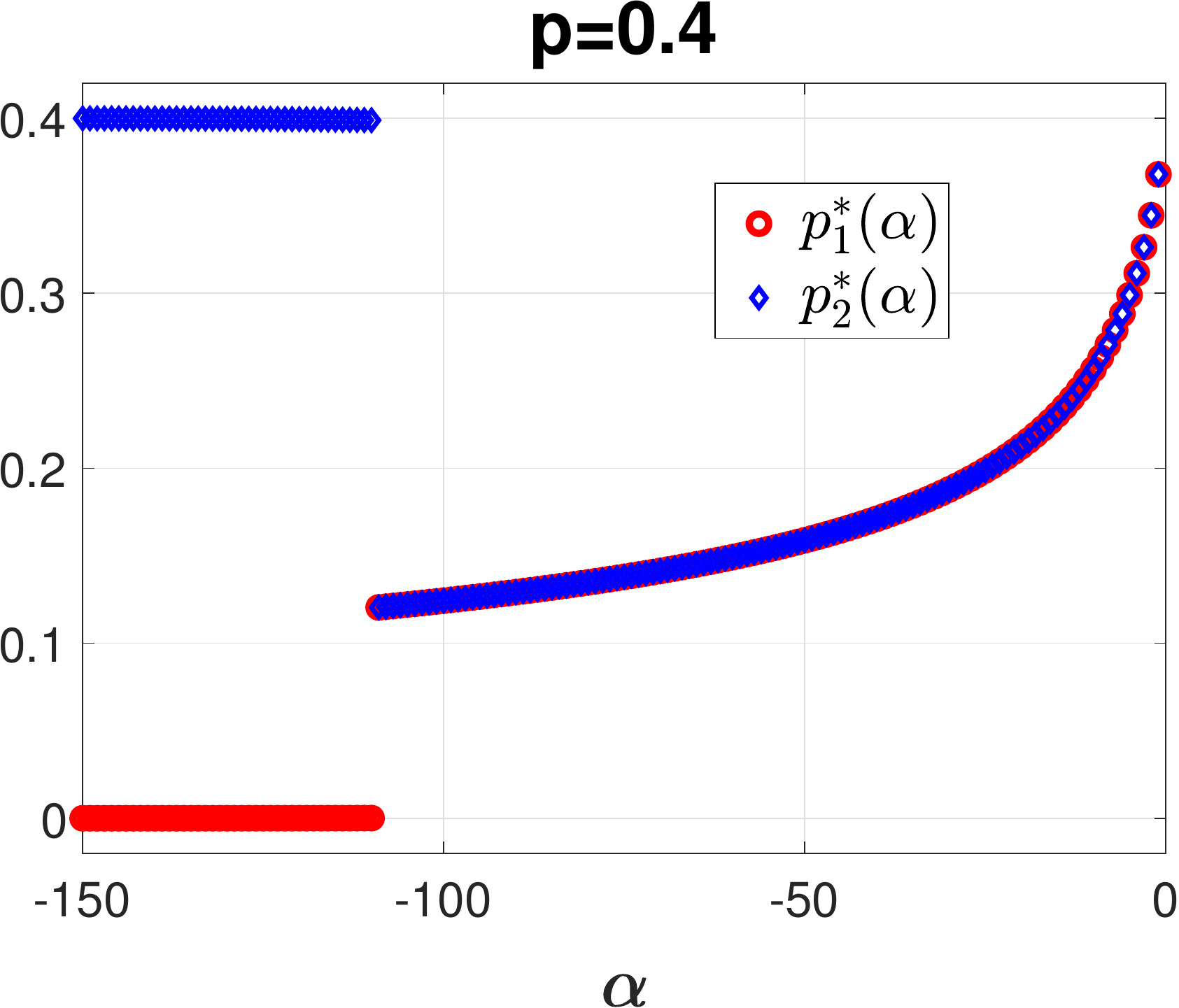}}
			\subfigure[\label{fig6b}]	
				{\includegraphics[scale=0.3]{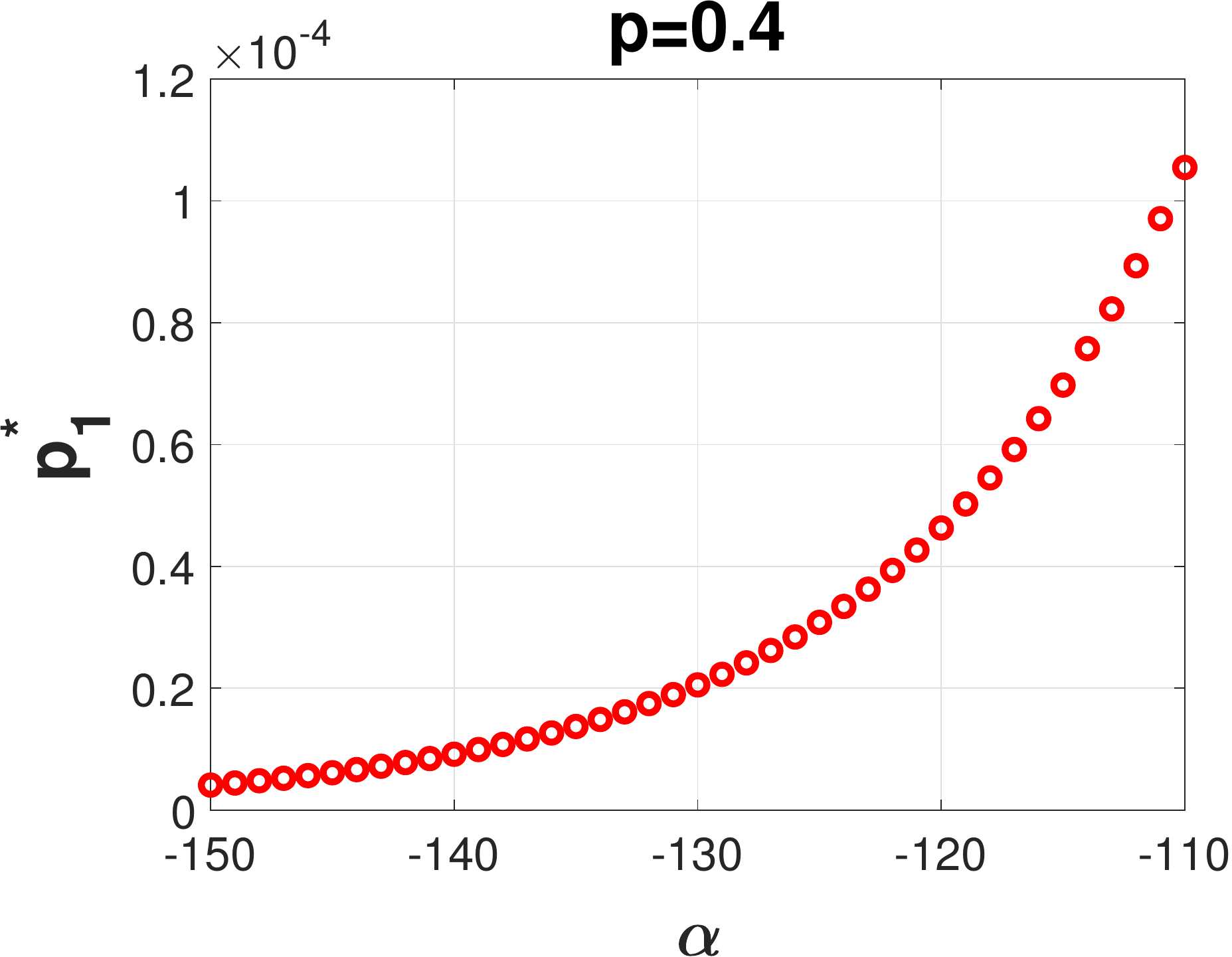}}
			\subfigure[\label{fig6c}]	
				{\includegraphics[scale=0.3]{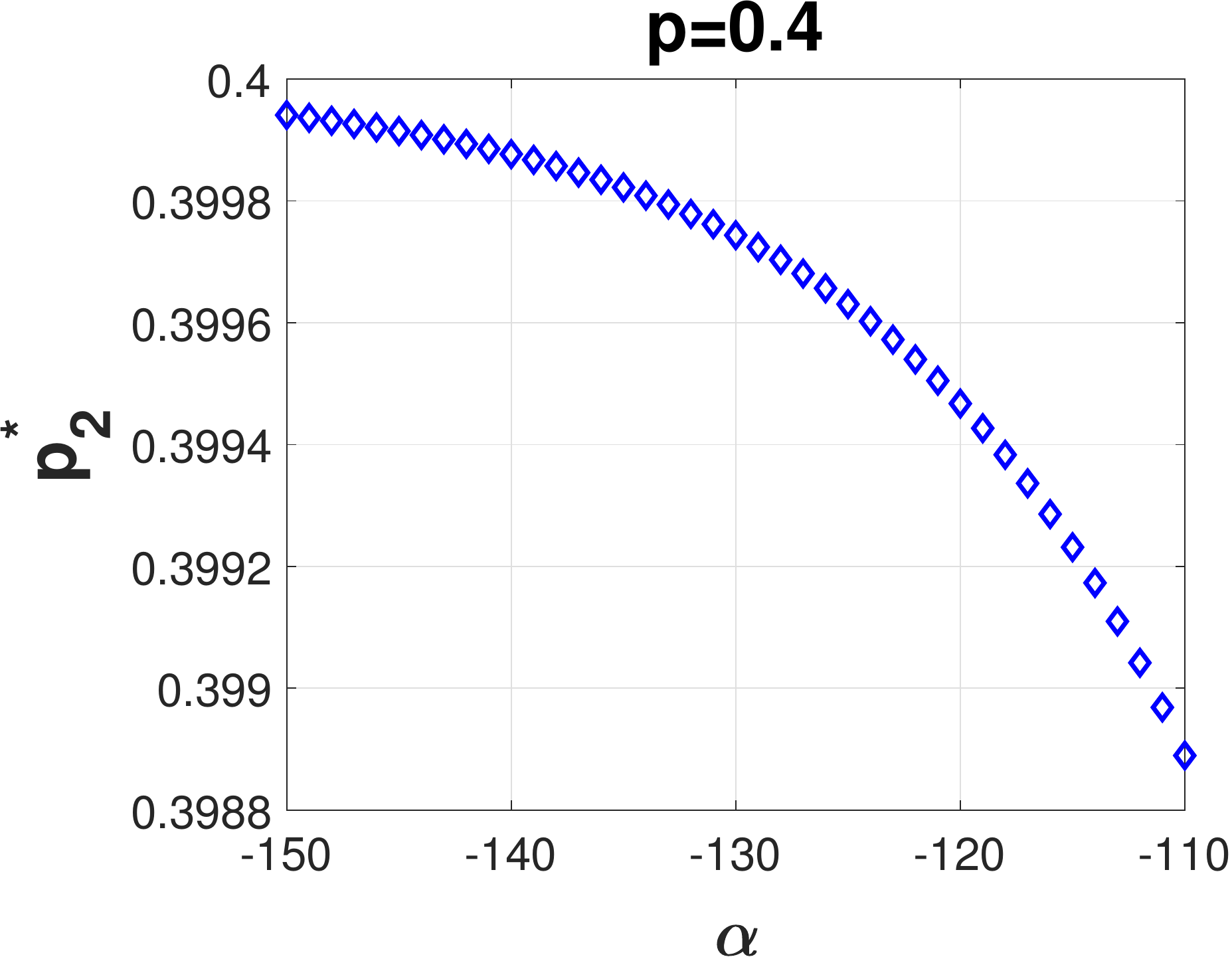}} \\
						\subfigure[\label{fig6d}]		
				{\includegraphics[scale=0.3]{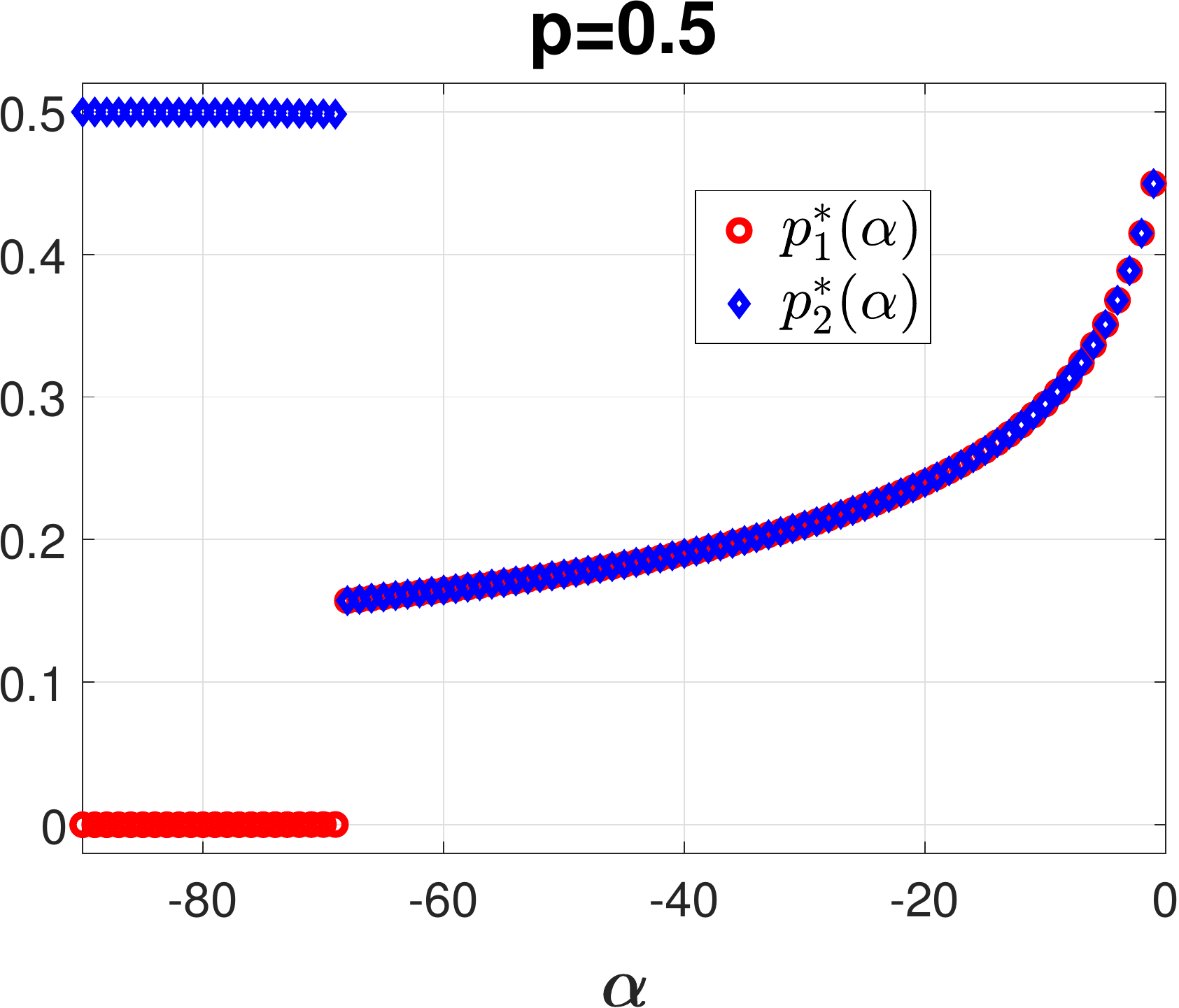}}
						\subfigure[\label{fig6e}]	
				{\includegraphics[scale=0.3]{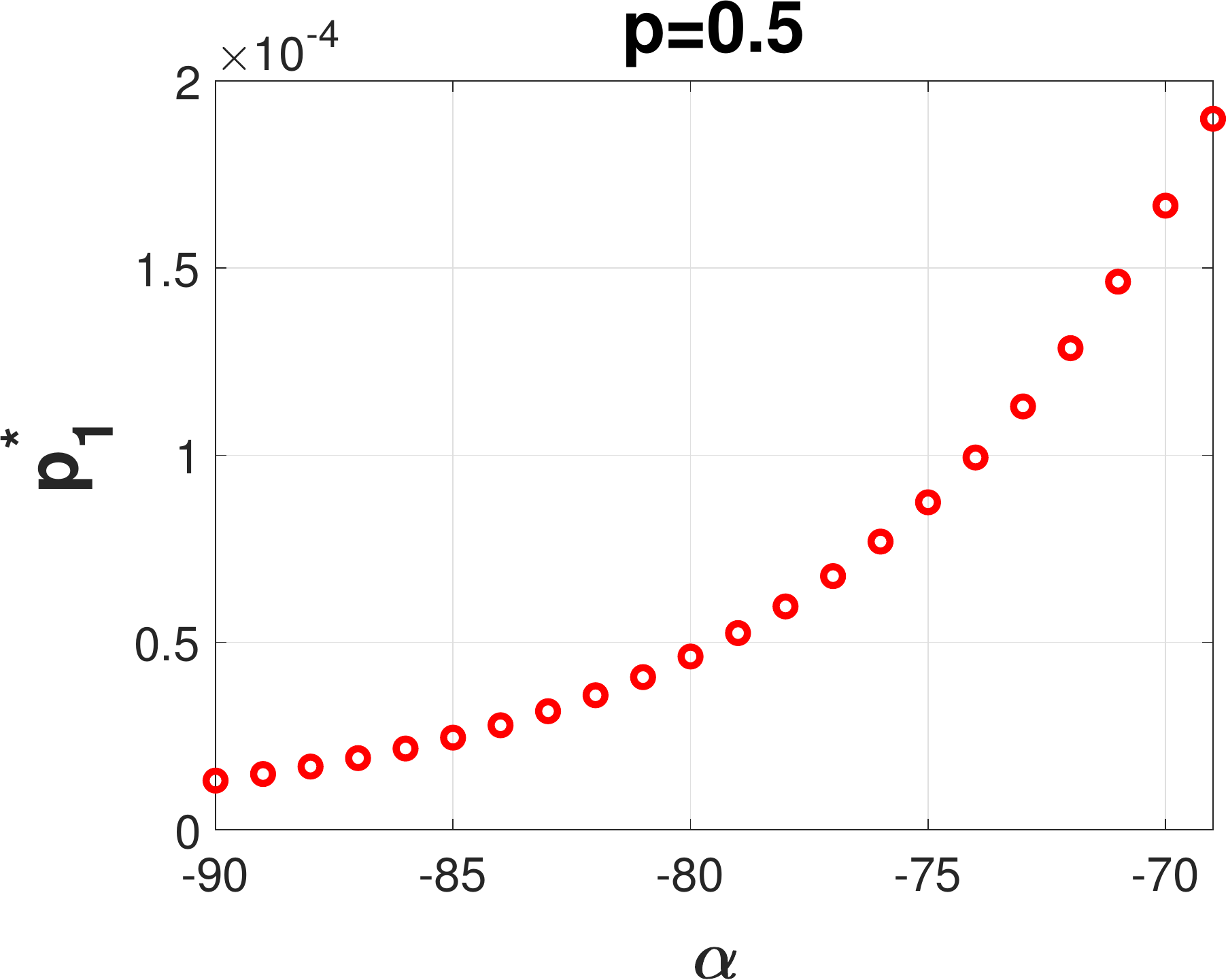}}
						\subfigure[\label{fig6f}]	
				{ \includegraphics[scale=0.3]{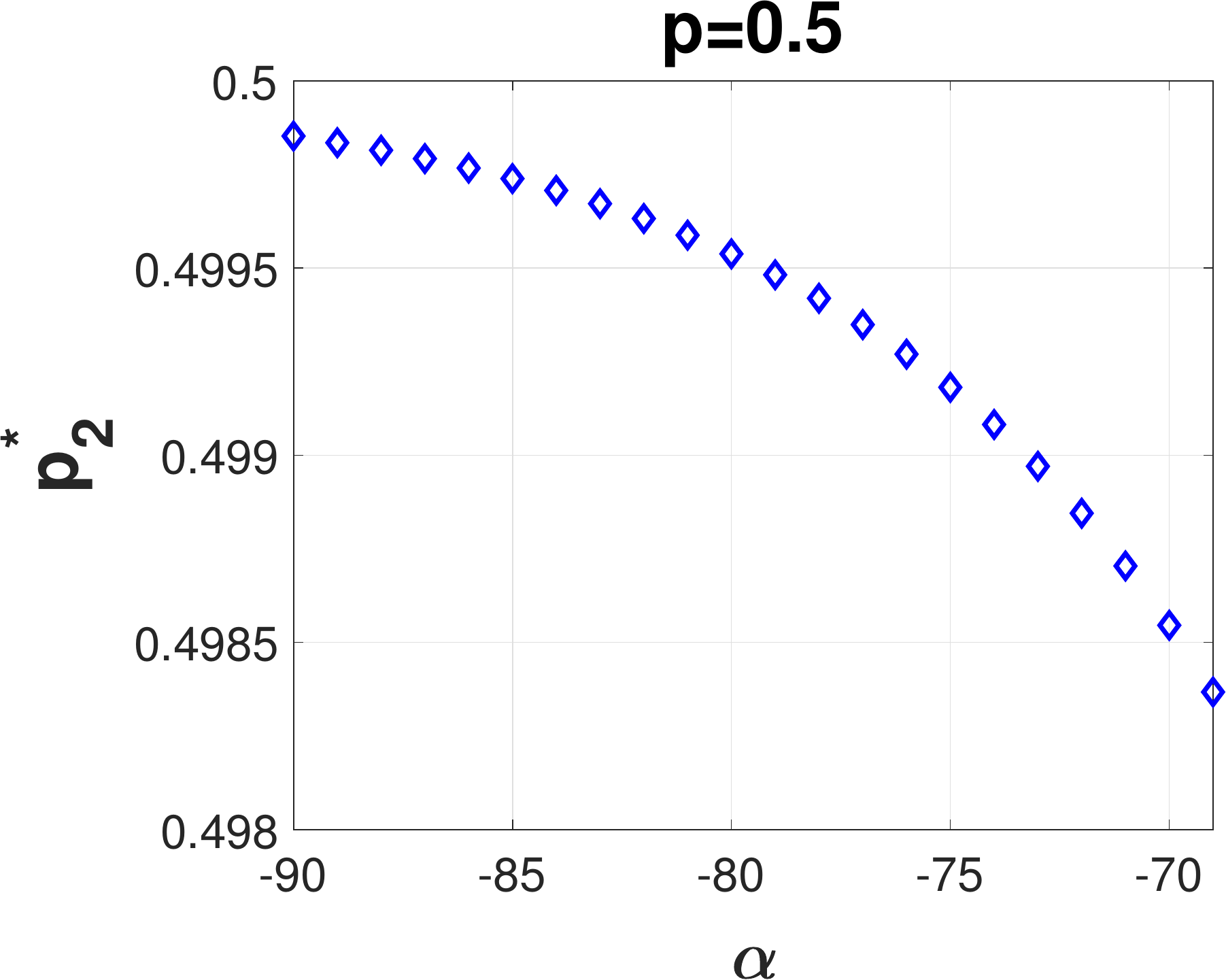}}
		\end{tabular}}
		\caption{Upper panel: the left picture represents the parameters  $p^*_{1}(\alpha)$ ($\circ$)  and 
		$p^*_{2}(\alpha)$  ($\diamondsuit$) of the optimizer of \eqref{funzionale}, as a function of  $\alpha\in\,[-150,0]$, where $\alpha$ varies with unitary step, and $p=0.4$. The pictures in the middle and in the right columns show the behaviour of $p^*_{1}(\alpha)$ and $p^*_{2}(\alpha)$ in the zoomed interval $[-150,-110]$. Lower panel: the same as in the upper panel for $p=0.5$,  $\alpha\in\,[-90,0]$ and $\alpha\in\,[-90,-69]$.}\label{graph_bip_optimizers}
	\end{center}
\end{figure}
\begin{figure}[h!]
	\begin{center}
		\makebox[\linewidth]{
			\begin{tabular}{ccc}
			\subfigure[\label{fig7a}]
				{\includegraphics[scale= 0.35]{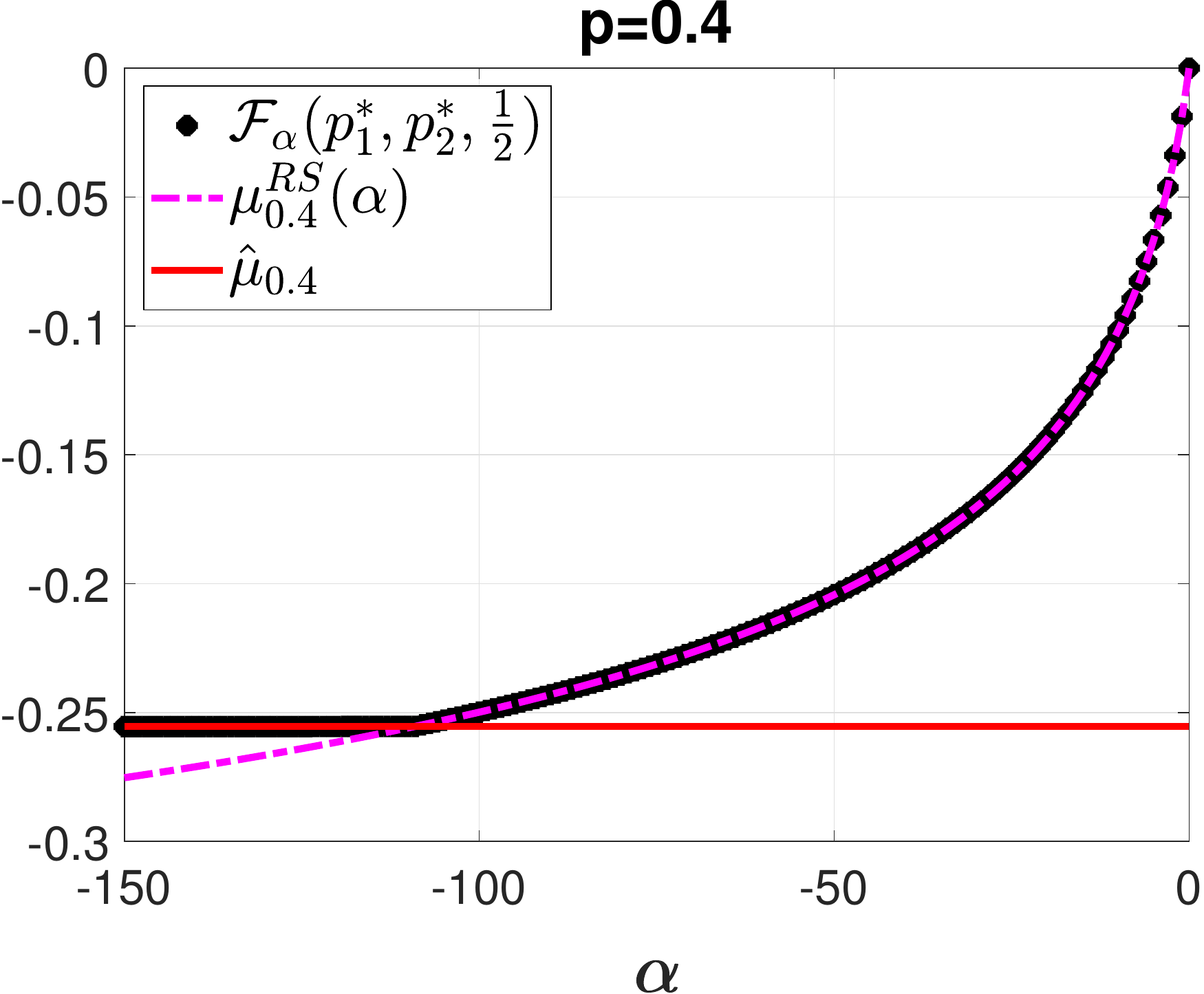}}
                \subfigure[\label{fig7b}]
				{\includegraphics[scale=0.35]{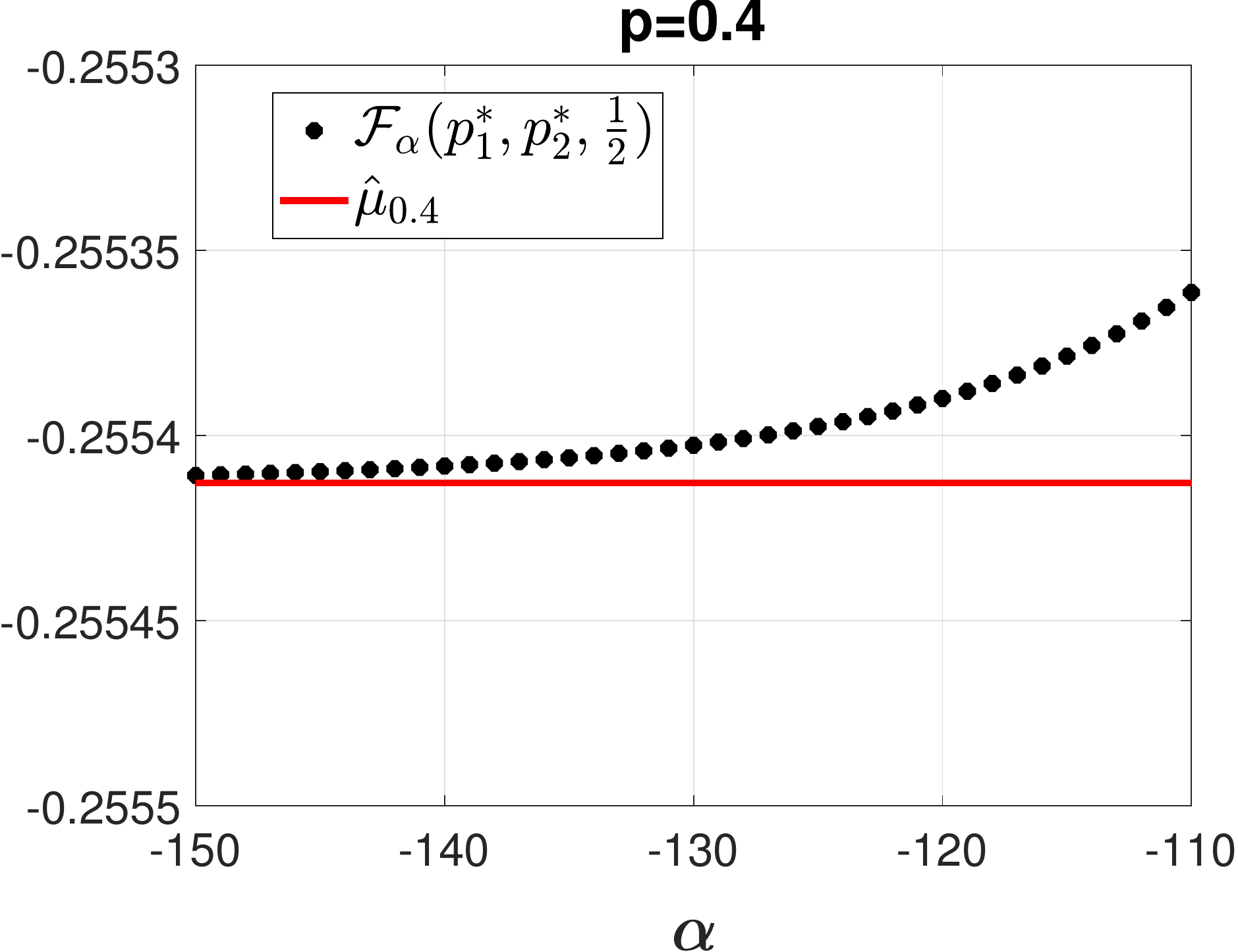}}
		\end{tabular}}
		\caption{Left column: plot of $\mathcal{F}_{\alpha}(p^*_{1}(\za),p^*_{2}(\za),\frac{1}{2})$ (black dots) together with the function $\mu^{RS}_p(\alpha)$ (magenta dashed-dotted line) for $p=0.4$. The horizontal red line represents the asymptotic value $\hat{\mu}_{p}$, see \eqref{chap1:largenegative_limit}. Right column: zoom on the interval $[-150,-110]$ for the case $p=0.4$.
		}\label{graph_bip_funzott}
	\end{center}
\end{figure}

\noindent
with $p_{1},p_{2},a\in[0,1]$ and $\alpha\leq\,-2.$

 It is  simple to check that the point $(p_{1}^*(\alpha),p_{2}^*(\alpha),a^*)$ with $p_{1}^*(\alpha)=p_{2}^*(\alpha)=u^*(\alpha)$ and any $a^*\in[0,1]$ is a solution to equation  \eqref{chap3:sistema_Grad} for any $\alpha$. Indeed, when $p_1=p_2$ equations (a) and (b) are equal and coincide with the fixed-point equation \eqref{eq_punto_fisso}.  From the numerical solution of  system \eqref{chap3:sistema_Grad}  as $\alpha$ is varied,  we got evidence of the existence of a threshold $\tilde{\alpha}_c({p})$ above which  $(u^*(\alpha), u^*(\alpha), a^*)$ is the maximum (this holds for any $a'$, being such parameter meaningless in the homogeneous case). Crossing $\tilde{\alpha}_c({p})$ from above,  a non-homogeneous solution with $a^*=1/2$ and $p^*_{1}(\alpha)\approx 0$,  $p^*_{2}(\alpha) \approx p$ arises; it turns out to be the maximum of the problem for $\alpha<\tilde{\alpha}_c({p})$. 
 
 We give more details on the procedure we followed in order to analyze the solutions  of \eqref{chap3:sistema_Grad}. First we observe that there are three special values of $a$, namely $a=0,\, \frac 1 2,\, 1$.  As we said above, 
 $a=0,\, 1$ correspond to the constant graphon that can be obtained also as a special case (with $p_1=p_2$) of  $a=\frac 12$. Thus  since the cases $a=0,\, 1$ can be absorbed in $a=\frac 1 2$,  we are left with the problem of analyzing  \eqref{chap3:sistema_Grad} for $a=\frac 1 2$ and $a\ne \frac 12$.
 
 \smallskip
 $\bullet${\bf\, Case $\mathbf a\ne \frac12$}. From equation (b) in  \eqref{chap3:sistema_Grad}  we get $p_1= \frac{1}{\alpha p_2} \ln\left(\frac{p_{2}(1-p)}{p(1-p_{2})} \right)$ that  replaced in (c) gives an equation for $p_2$:
 \be\label{equazg}
 \kG_\alpha(p_2):=\frac{1}{\alpha^2 p_2^3} \left [\fpp \right ]^3 - p_2 \fpp + 2 I_p(p_2) -2 I_p\left (\frac{1}{\alpha p_2}\fpp \right)=0.
 \ee   
We solved numerically the previous equation for several values of $p$ and $\alpha$. The function $\kG_\alpha(p_2)$ turns out to be a concave function whose maximun, that is also  the solution to eq.\eqref{equazg}, coincides with the solution of eq. \eqref{eq_punto_fisso}, i.e. $p_2=u^* (\alpha)$, see  Fig.\ref{nuovafigura}. Thus from  eq. (b) we get a linear equation for $p_1$: 
$$
\alpha\, p_1 u^*(\alpha)= \ln\left(\frac{u^*(\alpha)(1-p)}{p(1-u^\star(\alpha))} \right),
$$
that has the solution $p_1=u^*(\alpha)$. Indeed, the equation $\alpha x^2 = \ln\left(\frac{x(1-p)}{p(1-x)} \right)$ is equivalent to the fixed-point equation  \eqref{eq_punto_fisso} that has an unique solution  for $\alpha < 0$.

From this discussion, we conclude that any maximizer with  $a\ne \frac 12$ is necessarily  the constant  graphon $( u^*(\alpha), u^* (\alpha), a)$. 

\begin{figure}[h!]
	\begin{center}
		\makebox[\linewidth]{
			\begin{tabular}{ccc}
				\includegraphics[scale= 0.6]{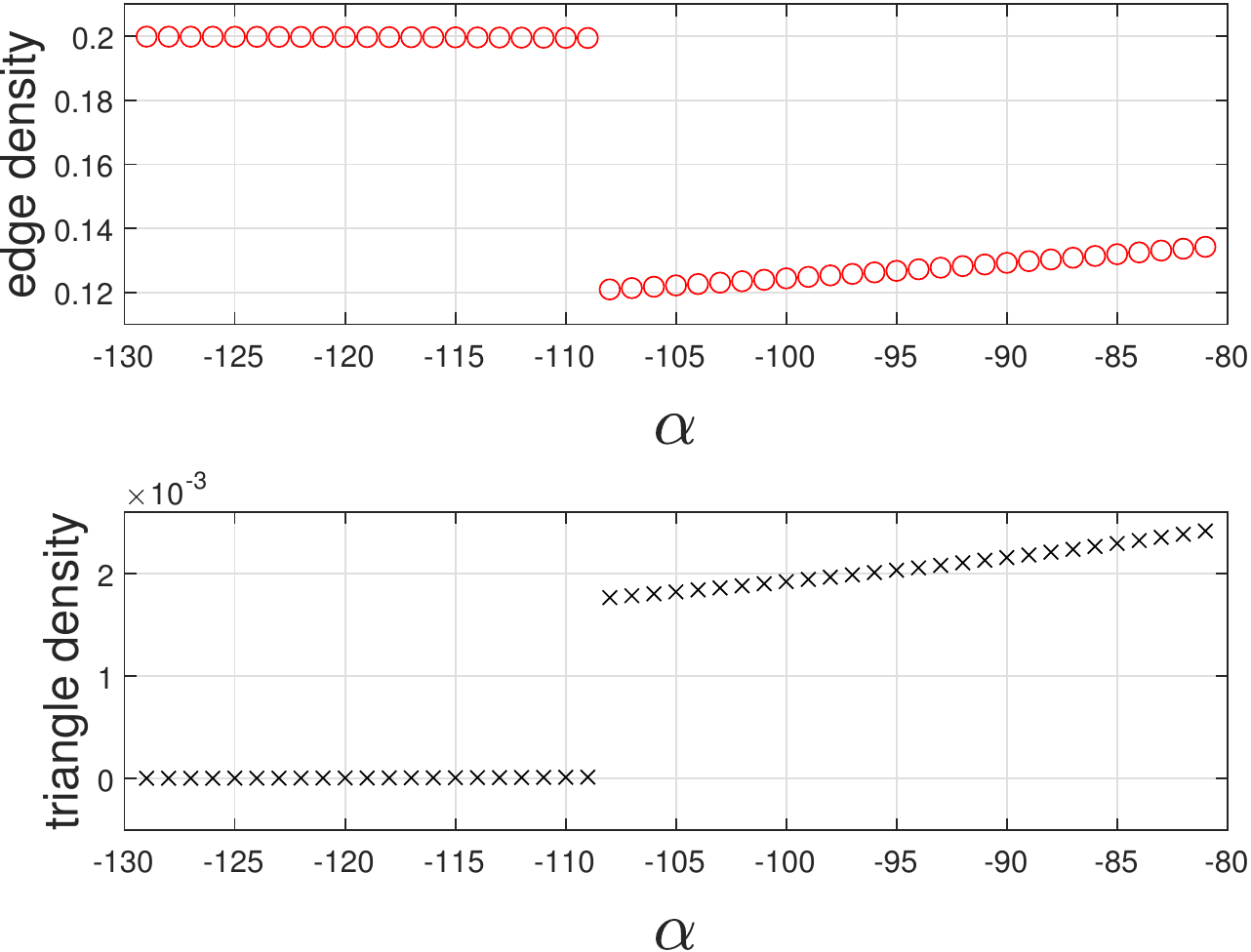}
			\end{tabular}}
		\caption{Density of edges and triangles of the optimizing graphon $g^{(\frac 1 2)}_{p_1^*(\alpha), p_2^*(\alpha)}$ of \eqref{funzionale}  for varying $\alpha$ and $a=\frac1 2$, $p=0.4$. The discontinuities make evident the phase transition at the critical point $\tilde{\alpha}_c(p)$. }\label{triang_edge}	
	\end{center}
\end{figure}

\smallskip
 $\bullet${\bf\, Case ${\mathbf a= \frac12}$}. In this case the function to be maximized takes the much simpler form $\kF_\alpha(p_1,p_2,\frac1 2)=\frac{\alpha}{12}(p_1^3+ 3 p_1p_2^2 )-\frac1 2 (I_p(p_1)+I_p(p_2) )$ with the stationarity condition:
 \begin{equation}\label{sistema_Grad_ridotto}
\begin{cases}
\frac{\alpha}{4} \left [  p_1^2+  p_2^2 \right ] -\frac1 2 \ln \left ( \frac{p_1(1-p)}{p(1-p_1)}\right )=0, &\quad (a) \\
p_{1}p_{2} -\ln\left(\frac{p_{2}(1-p)}{p(1-p_{2})} \right)=0. &\quad (b)\\
\end{cases} 
\end{equation}
%
We have computed the numerical solution $(p_1^*(\alpha), p_2^* (\alpha) )$ for several values of $p$, as in Fig.\ref{graph_bip_optimizers}, \col{finding a discontonuos transition between two solutions}.
More precisely, the left panels show that crossing a critical value  $\widetilde{\za}_c(p)$ from above  the transition between the constant graphon ($p_1^*(\alpha)= p_2^* (\alpha) )$) and a 1-step replica symmetry breaking  regime takes place. In particular, at $\widetilde{\za}_c(p)$ the solution jumps towards the   point $(0,p)$.
The central and right columns of Fig.\ref{graph_bip_optimizers} represent the solution  below the critical value showing that, in the limit  $\alpha \to -\infty$, the values $p^*_1(\za)$ and $p^*_2(\za)$  converge to the limits $0$ and $p$,   respectively, as conjectured in \eqref{eq:limpi} and \eqref{eq:limpb}.
A further evidence of this transition is given in Fig.\ref{graph_bip_funzott} in which the curves $\mathcal{F}_{\alpha}\left(p^*_{1}(\za),p^*_{2}(\za),\frac{1}{2}\right)$ and  
$\mu^{RS}_{p}(\alpha)$
are displayed. Above the critical value  $\widetilde{\alpha}_{c}({p})$ the two curves overlap thus revealing the replica symmetric phase whereas they separate below (replica breaking regime). 
For $\alpha < \tilde{\alpha}_c(p)$,  the right panel Fig.\ref{graph_bip_funzott}  shows that
$\mathcal{F}_{\alpha}\left(p^*_{1}(\za),p^*_{2}(\za),\frac{1}{2}\right)$ is
very close to the the asymptotic value 
\eqref{chap1:largenegative_limit}
(the discrepancy vanishes as $\alpha\to -\infty$).
\cla{Figure \ref{triang_edge} represents  the density of triangles and edges corresponding to the  maximizer $g^{(\frac 1 2)}_{p_1^*(\alpha), p_2^*(\alpha)}$. The former quantity is  given in \eqref{triangrsb} while the latter  is
$$ e(g^{(a)}_{p_{1},p_{2}}) = 
\int_{[0,1]^2} g^{(a)}_{p_{1},p_{2}} (x,y)\, dx\, dy = p_{1}[a^{2}+(1-a)^2] + 2 p_2 a (1-a).$$
The jump discontinuity shown in Fig.\ref{triang_edge}, located at $\tilde{\alpha}_c(p)$, makes evident the existence of the transition that separates the replica symmetric phase, in which both quantities decrease for decreasing $\alpha$, from the replica breaking phase. In this phase the density of edges gets close to the value $\frac p 2$, i.e. the density of the limiting equipartite  graphon \eqref{chap1:def:graphon_bip}, while the density of triangle  jumps towards a value close to zero, being zero the density of triangles of the same graphon.  }

\begin{figure}[htpb!]
	\begin{center}
		\makebox[\linewidth]{
			\begin{tabular}{cc}
				\includegraphics[scale=0.4]{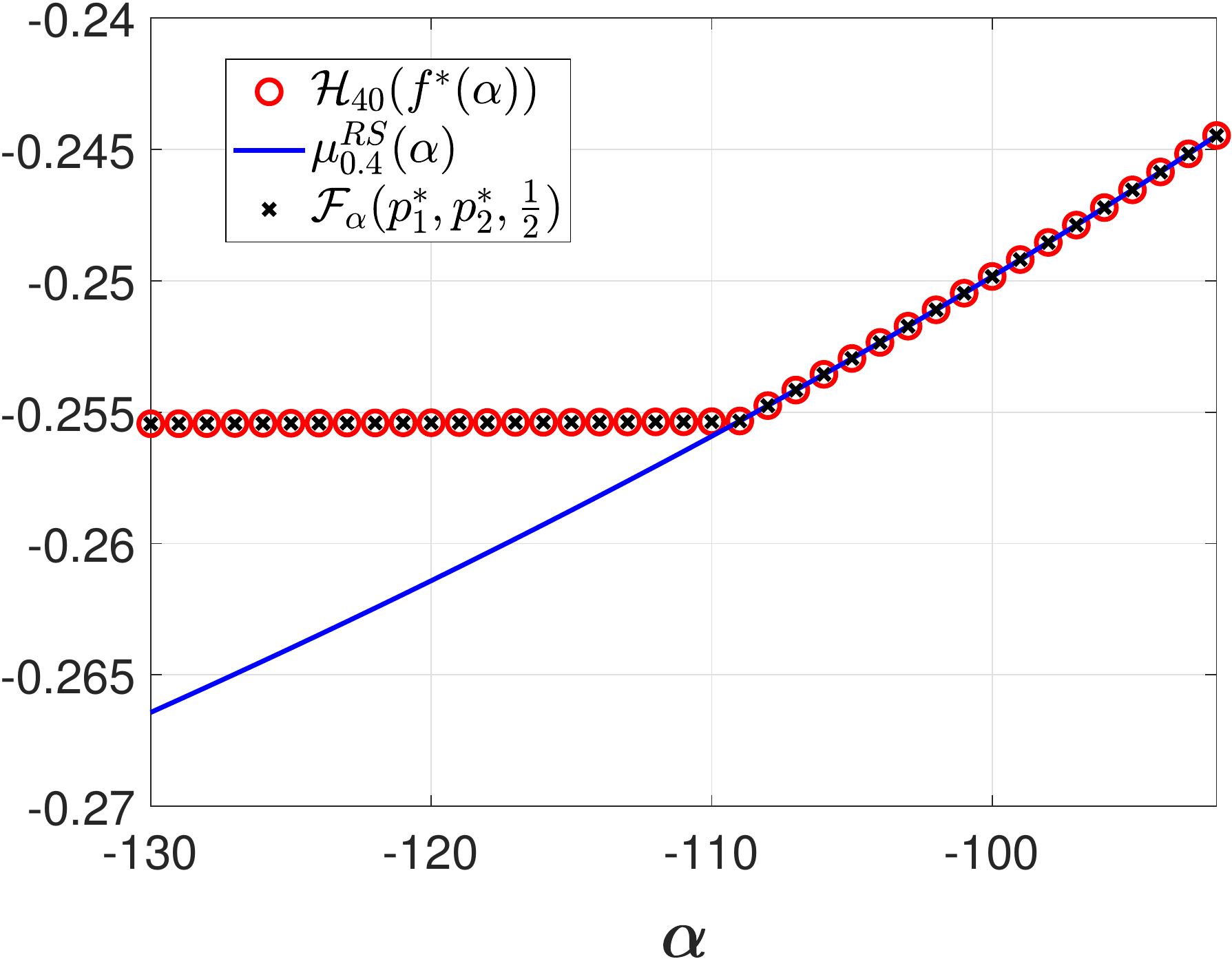}	
		\end{tabular}}
		\caption{The plot represents  $\mu^{1 RSB}_p(\alpha)=\mathcal{F}_{\alpha}\left(p^*_{1}(\za),p^*_{2}(\za),\frac{1}{2}\right) $ (black crosses), the solution  ${\cal H}_{40}(f^*(\alpha))$ of the discretized problem  \eqref{probl_v_distcretizato} (red circles) together with  $\mu^{RS}_p(\alpha)$. The picture refers to $p=0.4$.}\label{fig:SovrappGP1Step}
	\end{center}
\end{figure}

We conclude our analysis by
Fig.\ref{fig:SovrappGP1Step}, which displays the cumulant generating function in replica symmetric regime \eqref{chap1:eq:probl_scalare},
its 1-step replica symmety breaking version \eqref{funzionale} and the cumulant generating function obtained
from the discretized variational problem. 
We can observe that the three curves perfectly match above the critical threshold $\tilde{\alpha}_c(p)$,
whereas in the subcritical phase the curve obtained from the solution
of the discretized problem agrees with the 1RSB  cumulant generating function and is
strictly larger than the replica symmetric  cumulant generating function.
Furthermore, Fig.\ref{fig:SovrappGP1Step} shows that $\tilde{\za}_c(p)$ 
is a singularity of $\za \to \mathcal{F}_{\alpha}(p^{*}_{i}(\za),p^{*}_{b}(\za),\frac{1}{2})$,
at which the left and right derivatives are different.

\section{Conclusion and perspectives}
\label{per}

\begin{figure}[h!]
	\begin{center}
		\makebox[\linewidth]{
			\begin{tabular}{ccc}
				\includegraphics[scale= 0.25]{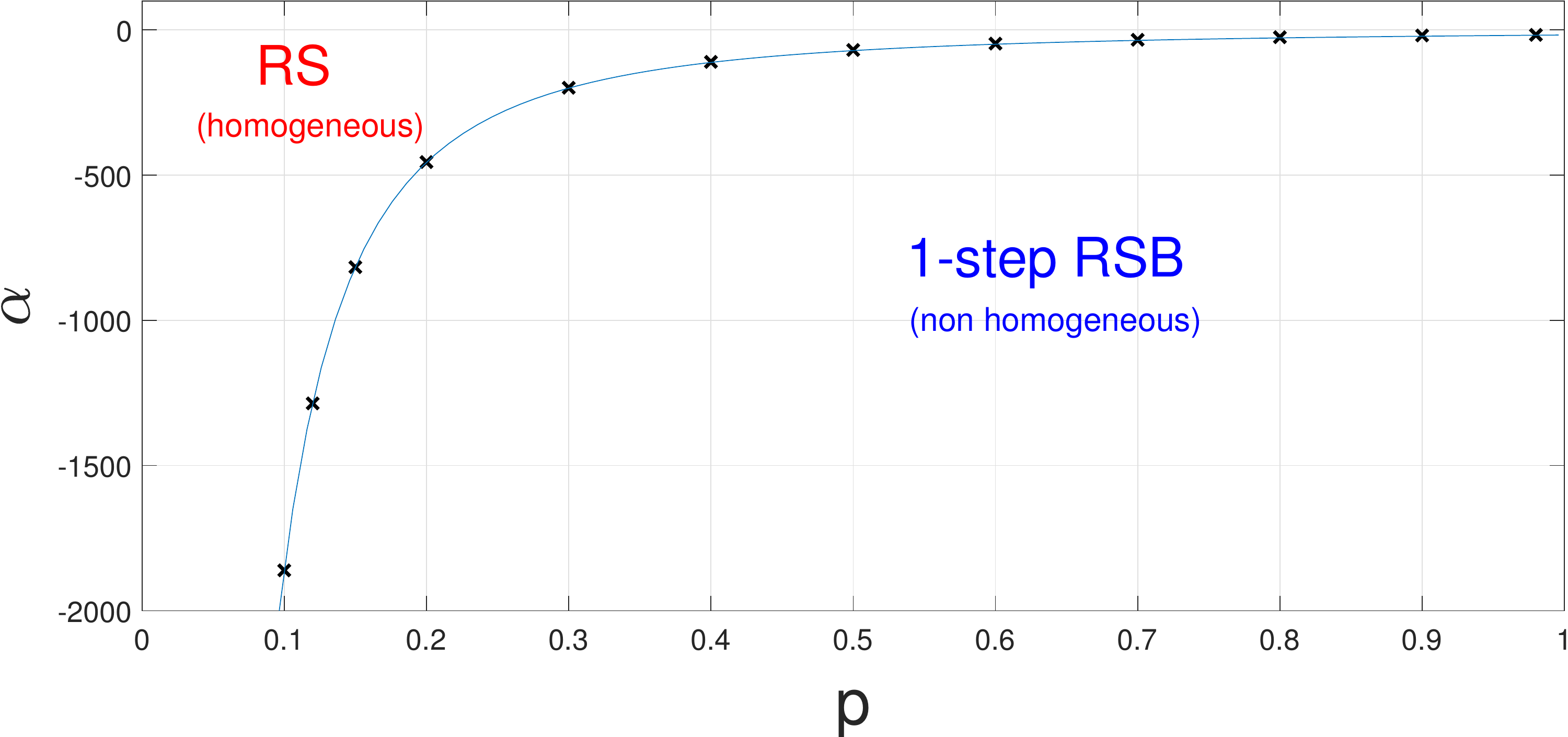}
			\end{tabular}}
		\caption{Phase diagram of the edge-triangle model in the $(\alpha,p)$ plane. We measured 
		critical points $\tilde{\alpha}_c(p)$ for several $p$ values. The method of least squares furnishes
		 the curve  $\tilde{\alpha}_{c}(p)=-17.66\, p^{-2.027}$, which is shown as a continuous line.}\label{curva_interpolante_tab}
				
	\end{center}
\end{figure}

\col{In this paper we identified a first order phase transition for the edge-triangle model.
The transition occurs between the replica symmetric regime,
where the typical graph is the constant graphon describing 
the \ER graph (i.e. independent edges, yet with a modified parameter for the probability of 
edges accounting for the imposed number of triangles) 
and the replica symmetric breaking regime,
where the typical graph is the graphon describing the 1-step replica symmetric broken solution 
(generalizing the bipartite random graphs). 
The phase diagram is cointained in Figure \ref{curva_interpolante_tab},  
showing the dependence of the critical value $\tilde{\alpha}_c(p)$ from $p$.
The data are in good agreement with a fit $ |\tilde{\alpha}_c(p)|\sim p^{-2}$.
As it is usual the case with first order phase transitions one might expect
metastability, however we did not investigate this in the present paper. 
}

\col{A final comment is in order on the problem of ensemble inequivalence.}
As the pressure $\psi_n(\beta_1,\beta_2)$ is the crucial quantity in the 
canonical ensemble, the entropy $s_n(e,t)$ defined as
\be
s_n(e,t) = \frac{1}{n^2}\ln \sum_{x \in \kX_{n}} \delta\left(E_n(x) - \frac{n^2 e}{2}\right) \delta\left(T_n(x) - \frac{n^3t}{6}\right) 
\ee
is the important quantity in the microcanical ensemble.
An heuristic application of the Laplace method would imply
\begin{eqnarray}
\psi(\beta_1,\beta_2) 
& = & 
\lim_{n\to\infty} \frac{1}{n^2} \ln \int  e^{n^2(\beta_1 e + \beta_2 t - s_n(e,t))} \,de \,dt \nonumber \\
& = & 
\sup_{e,t} \Big( \beta_1 e + \beta_2 t - \lim_{n\to\infty} s_n(e,t)\Big) \nonumber \\
& = & 
\sup_{e,t} \Big( \beta_1 e + \beta_2 t - s(e,t)\Big), 
\end{eqnarray}
i.e. in the thermodynamic limit the canonical pressure can be obtained
from the microcanonical entropy by a Legendre transform. 
One then says that the two ensembles are {\em thermodynamic equivalent} if 
such correspondence also holds in the reversed direction.
This is the same as requiring that the microcanonical entropy
is strictly convex, i.e. the involution property of the Legendre 
transform for strictly convex functions.
%
%
%
%
%

\cla{The problem of {\em ensemble inequivalence} is, in fact,
more general than just thermodynamic inequivalence
 (see \cite{touchette2015equivalence}
for a recent account).}
When the correspondence via Legendre transform between pressure
and entropy does not hold, the difference between canonical 
and microcanonical ensemble can then be probed in several ways.
In this paper we have focused on {\em macrostate inequivalence}.
Namely, we asked if sampling a very large 
graph uniformly at random from the set of all graphs with given dentities
of edges and triangles $(e^*,t^*)$ 
is statistically equivalent to sampling  a very large graph from the 
edge-triangle model with parameter values $\beta_1(e^*,t^*),\beta_2(e^*,t^*)$ 
that are obtained by inverting the relations \eqref{averages2} 
with $\langle e \rangle = e^* $ and  $\langle t \rangle = t^*$. 
As we have discussed, in the thermodynamic limit  
graphs are described by the notion of
graphon, thus the macrostate inequivalence
amounts to ask if the two sampling procedures
produce different graphons.

{The sampling from the microcanonical ensemble has been investigated for instance in \cite{pikhurko2017asymptotic,radin2013phaseCN,radin2015singularities,kenyon2017multipodal,aristoff2015asymptotic}.}
It has been found that the structure of graphs drawn from the microcanonical ensemble 
is very rich and may vary a lot as a function of the number of prescribed edges and triangles.
For instance, for a choice $(e,t)$ such that $e=\frac12 + \epsilon$ with $\epsilon \in (\frac{l-2}{2l},\frac{l-1}{2l+2})$ with $l\in\N\setminus \{1\}$ and $t$ on the scallopy curve,
the vertex set of a graph drawn from the microcanonical ensemble can be partitioned into $\ell$ subsets
($\ell-1$ of them of the same size and the last of different size).
The graph has the form of a complete $\ell$-partite graph on these pieces, plus some additional edges in the last 
piece that create no additional triangles \cite{den2018ensemble}.

\col{
In contrast to the microcanonical sampling, our numerical analysis suggests that 
in the description of the canonical sampling no higher level of replica symmetry breaking is required.
Our results imply that the ensemble inequivalence  found in \cite{den2018ensemble}
by  a positive relative entropy between the microcanonical and canonical measures,
does indeed carry over to macrostate inequivalence. 
So far, no signatures of the first order phase transition we observed in the canonical
ensemble has been found in the microcanonical ensemble.}

\section*{Acknowledgments} 
We acknowledge an enlightening discussion with Remco van der Hofstad.  
We thank M. Prato and S. Rebegoldi  for making available to us their Gradient Projection code.
\section*{Appendix: $n=3$}
\begin{figure}[htpb!]
	\begin{center}
		\makebox[\linewidth]{
			\begin{tabular}{cc}
				\includegraphics[scale=0.45]{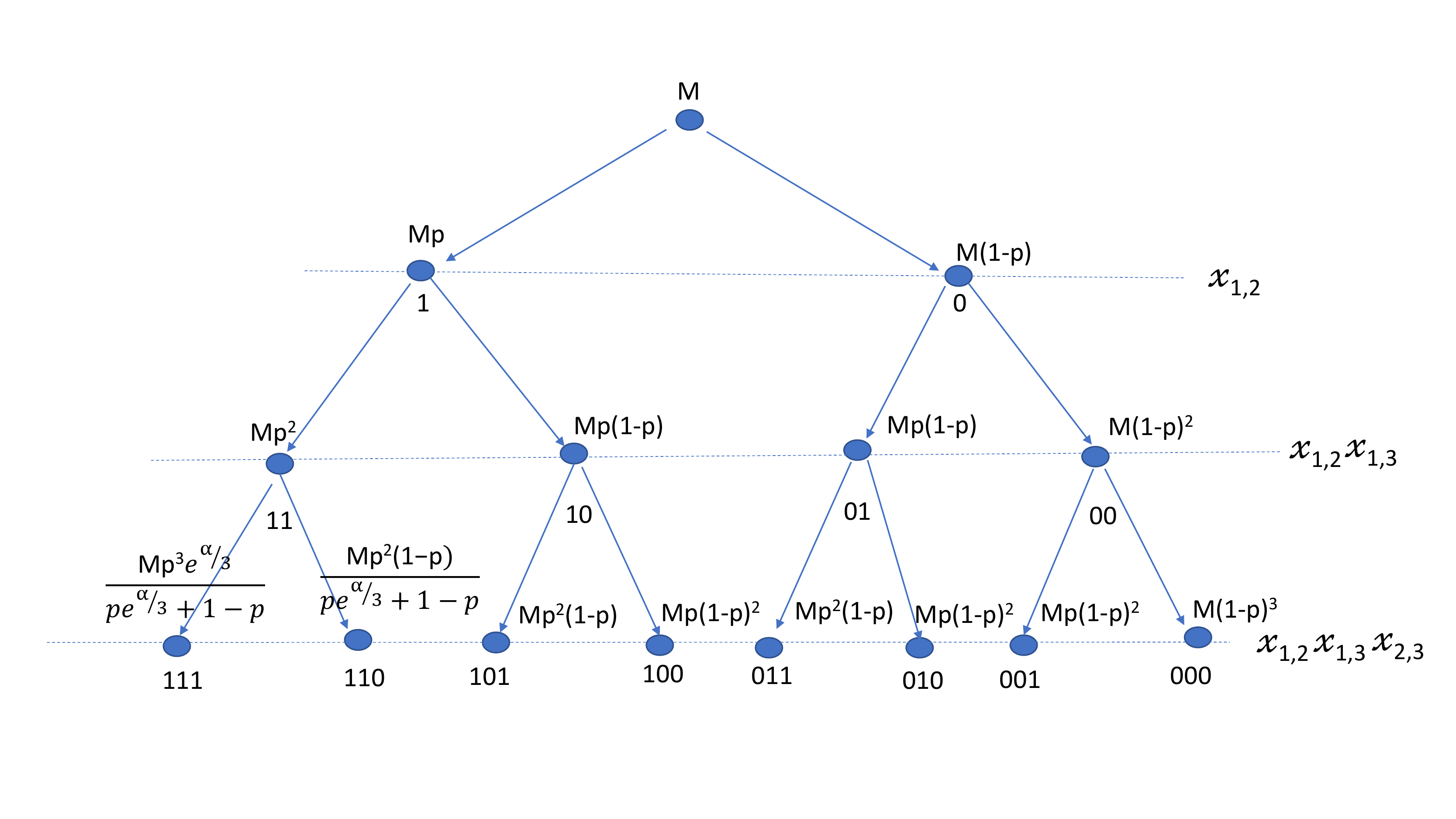}	
		\end{tabular}}
		\caption{Graphical representation of the evolution step, according to the tilted dynamics $P_\alpha(\cdot, \cdot)$ of cloning scheme. The levels of the tree display
		the values of the elements of the adjacency matrix $x_{1,2}, x_{1,3}, x_{2,3}$ during the evolution leading to $G_3$. $M$ is the initial size of the population. In all levels the expected  sizes of the 
		sub-polulations with a given configuration of values  $x_{1,2}, x_{1,3}, x_{2,3}$ are reported.}\label{fig:albero}
	\end{center}
\end{figure}

In order to show how the cloning algorithm works, we explicitly compute the cumulant generating function of triangles in the simplest case, that is the graph of size $n=3$. 
Since the probability of  the unique triangle is $p^3$, we have 
\be\label{eq:mu3}
\mu_{3,p}(\za) = \frac 1 3 \ln \left \langle \exp \frac{\alpha}{3} T_3(X) \right \rangle^{ER}_3 = \frac 1 3 \ln   \left [\e^\frac{\za}{3}p^3 +1-p^3 \right] .
\ee
We compute again \eqref{eq:mu3} by applying the dynamics described in Sec.\ref{sectCloningRG} to a family of $M$ clones. The three steps required to construct the edges of the graph $G_3$ are represented in Fig.\ref{fig:albero}. 
The leafs of the tree represent the occupation variables of the three possible edges, $x_{1,2}, x_{1,3}, x_{2,3}$. 
In the first step the edge connecting two vertices, say 1 and 2, of each clone is added with probability $p$. Thus, in the clone population about $pM$ graphs have the edge $(1,2)$  and $(1-p)M$ graphs have not this edge, see the first level in Fig.\ref{fig:albero} . In the second step the edge $(1,3)$  is added, still with probability $p$, leading to four possible values for the pair $(x_{1,2},x_{1,3})$. 
Thus, the expected numbers of types  are $Mp^2, Mp(1-p), Mp(1-p), M(1-p)^2$, see level 2 in Fig.\ref{fig:albero}. Let us observe that in the first two steps, since $\Delta T(x,y)=0$, the original and tilted transition probabilities coincide:  $P_\za(x,y)=P(x,y)$, see \eqref{eq:palfat}.
The situation changes in the last step. Indeed, the configuration of edges $x=11$ (which means that $x_{1,2}=1$ and $x_{1,3}=1$) may evolve to $y=111$, with $\Delta T(x,y)=1$, or to $y=110$, in which case $\Delta T(x,y)=0$. For all the other configurations $x$ we have $\Delta T(x,y)=0$. Thus, see the definition \eqref{eq:palfat} of the tilted probability:
\be
\label{pierho}
P_\za(11,111)= \frac{\e^\frac{\za}{3}p}{ \e^\frac{\za}{3}p+1-p}, \quad  P_\za(11,110)= \frac{1-p}{ \e^\frac{\za}{3}p+1-p},
\ee
being $P(11,111)=p$, $P(11,110)=1-p$ and $k_\za(11)= p\e^\frac{\za}{3}+1-p$. Then, the probability of the paths connecting the root $\phi$ to the leafs  $111$, respectively $110$, will be, :
\be
\pp(\phi,111)= \frac{\e^\frac{\za}{3}p^3}{ \e^\frac{\za}{3}p+1-p},\quad \pp(\phi,110)= \frac{p^2(1-p)}{ \e^\frac{\za}{3}p+1-p}.
\ee
The average  in \eqref{eq:mut2t} can be computed as the sum over the paths from the root to the leafs  (equivalently, as the sum over the leafs).  Recalling that   $k_\za(11)=  p\e^\frac{\za}{3}+1-p $ and observing also that $k_\za(x)=1$ if $x\ne 11$, from \eqref{eq:mut2t}  we have:
\eqan{
	\mu_{3,p}(\za) 
	&= \frac 1 3 \ln  \left [\pp(\phi,111) k_\za(11)+ \pp(\phi,110) k_\za(11)+ \sum_{x\notin \{111, 110\}}\pp (\phi,x )   \right ] \label{eq:mu3a}\\
	&= \frac 1 3 \ln  \left [\frac{p^3\e^\frac{\za}{3}}{ p\e^\frac{\za}{3}+1-p} (p\e^\frac{\za}{3}+1-p) +  \frac{p^2(1-p)}{ p\e^\frac{\za}{3}+1-p} (p\e^\frac{\za}{3}+1-p) + (1-p^2) \right ]\\
	&=  \frac 1 3 \ln \left [ p^3\e^\frac{\za}{3} +1-p^3 \right ],\label{eq:mu3b}
}
The cloning algorithm simulates \eqref{eq:mu3b} by producing a final population of expected size 
$$
M_3=M\,\pp(\phi,111) k_\za(11)+ M\, \pp(\phi,110) k_\za(11)+\sum_{x\notin \{111, 110 \}} M\, \pp (\phi, x).
$$
Then, the cumulant generating function is computed from the size of the final population $M_3$ as
\be
\mu_{3,p}(\za) =  \frac 1 3 \ln   \left [ \frac{M_3}{M} \right ] .
\ee

\end{document}